\documentclass[preprint,secnumarabic,showpacs,amsmath,amssymb,floatfix,superscriptaddress,longbibliography,aps,prb]{revtex4-1}

\usepackage{graphicx}% bilder einbinden  %% [demo] entfernt die bilder
\usepackage{xcolor}%
\usepackage{bm}% bold math
\usepackage{paralist}
\usepackage[utf8]{inputenc}
\newlength{\alphabet}
\usepackage{hyperref} % links in text
\usepackage{verbatim} % mehrzeilige kommentare
\usepackage{amsmath} % mathe power ka was was macht
\usepackage{amsfonts}
\usepackage{txfonts}
\usepackage{amssymb}
\usepackage{tabularx}  % für tabelle bis zum rand: das X nicht vergessen
\usepackage{units} % für \nicefrac hübsche schiefe bruchstriche
\usepackage{physics} % \bra \ket \expval{}{}
\usepackage{bbold} % hübscher einsoperator \mathbb{1}
\usepackage{slashed}
\usepackage{float}
\usepackage[english]{babel}
\usepackage{caption}
\newcommand{\lhf}{LiHoF$_{4}$}

\newcommand{\vcf}{V_{\rm CF}}
\newcommand{\bc}{$B_{\rm c}$}
\newcommand{\tc}{$T_{\rm c}$}

\renewcommand{\vec}[1]{{\bf #1}}

\newcommand{\be}{\begin{equation}}
\newcommand{\ee}{\end{equation}}
\begin{document}
%%%%%%%%%%%%%%%%%%%%%%%%%%%%%%%%%%%

\renewcommand{\floatpagefraction}{0.5}
\bibliographystyle{nature}

%%%%%%%%%%%%%%%%%%%%%%%%%%%%%%%%%%%

\title{Emergence of mesoscale quantum phase transitions in a ferromagnet}

%%%%%%%%%%%%%%%%%%%%%%%%%%%%%%%%%%%

\author{Andreas Wendl\footnote{These authors contributed equally.}}
\affiliation{Physik Department, Technische Universit\"at M\"unchen, 85748 Garching, Germany}

\author{Heike Eisenlohr$^{\rm a}$}
\affiliation{Institut f\"ur Theoretische Physik and W\"urzburg-Dresden Cluster of Excellence ct.qmat, Technische Universit\"at Dresden, 01062 Dresden, Germany}

\author{Felix Rucker}
\affiliation{Physik Department, Technische Universit\"at M\"unchen, 85748 Garching, Germany}

\author{Christopher Duvinage}
\affiliation{Physik Department, Technische Universit\"at M\"unchen, 85748 Garching, Germany}

\author{Markus Kleinhans}
\affiliation{Physik Department, Technische Universit\"at M\"unchen, 85748 Garching, Germany}

\author{Matthias Vojta\footnote{matthias.vojta@tu-dresden.de}}
\affiliation{Institut f\"ur Theoretische Physik and W\"urzburg-Dresden Cluster of Excellence ct.qmat, Technische Universit\"at Dresden,
01062 Dresden, Germany}

\author{Christian Pfleiderer\footnote{christian.pfleiderer@tum.de}}
\affiliation{Physik Department, Technische Universit\"at M\"unchen, 85748 Garching, Germany}
\affiliation{Center for QuantumEngineering (ZQE), Technische Universit\"at M\"unchen, 85748 Garching, Germany}
\affiliation{Munich Center for Quantum Science and Technology (MCQST), Technische Universit\"at M\"unchen, 85748 Garching, Germany}

\date{\today}

\maketitle

%%%%%%%%%%%%%%%%%%%%%%%%%%%%%%%%%%

\textbf{
Mesoscale patterns as observed, e.g., in ferromagnets, ferroelectrics, superconductors, mono-molecular films, or block-copolymers \cite{1995_Seul_Science,2008_Hubert_Book}, reflect spatial variations of a pertinent order parameter at length- and time-scales that may be described classically. This raises the question for the relevance of mesoscale patterns near zero temperature phase transitions, also known as quantum phase transitions (QPTs).
Here we report the magnetic susceptibility of {\lhf} -- a dipolar Ising ferromagnet -- near a well-understood transverse-field quantum critical point (TF-QCP) \cite{1996_Bitko_PRL, 2011_Sachdev_Book}. When tilting the magnetic field away from the hard axis such that the Ising symmetry is always broken, a line of well-defined phase transitions emerges from the TF-QCP characteristic of an additional symmetry breaking, in stark contrast to a crossover expected microscopically.
We show that a continuous suppression of ferromagnetic domains, representing a breaking of translation symmetry on mesoscopic scales in an environment of broken magnetic Ising symmetry on microscopic scales, is in excellent qualitative and quantitative agreement with the field- and temperature dependence of the susceptibility and the magnetic phase diagram of {\lhf} under tilted field.
This identifies a new type of phase transition that may be referred to as mesocale quantum criticality, which emanates from the text-book example of a microscopic ferromagnetic TF-QCP. Our results establish the surroundings of QPTs as a regime of mesoscale pattern formation, where non-analytical quantum dynamics and materials properties without classical analogue may be expected.
}
%\\\\
%\noindent [abstract: maximum 200 words / present version: 269 words]

%%%%%%%%%%%%%%%%%%%%%%%%%%%%%%%%%%
\newpage

%\section{Mesoscale textures and key message}

In recent decades a plethora of new phenomena reflecting strong quantum correlations have been discovered \cite{2008_Giamarchi_NatPhys, 2004_Senthil_Science, 2017_Zhou_RMP, 2017_Wen_RMP, 2007_HvL_RMP}.
Remarkably, many of them appear to involve mesoscale patterns, with key examples including the interplay of ferromagnetic and antiferromagnetic domains \cite{2020_Scheie_PNAS}, the condensation of topological defects in valence bond solids \cite{2004_Senthil_Science}, or the formation of spin-density wave order at ferromagnetic quantum criticality \cite{2009_Conduit_PRL, 2015_Abdul_NatPhys}.
Yet, starting purely from microscopic considerations, an account of mesoscale textures represents a major challenge. Moreover, such textures are typically described at a classical level.
To address the emergence of textures in the presence of well-understood quantum fluctuations we studied the arguably simplest example of a quantum phase transition\cite{2007_HvL_RMP}, namely the response of an insulating easy-axis ferromagnet to a magnetic field, $B$, applied transverse to the preferred magnetization axis. This transverse-field quantum magnetism is of broad interest, as it allows to benchmark the entanglement of complex spin systems, the tunnelling of single moments and domain walls, quantum annealing, and Rabi oscillations  \cite{2011_Sachdev_Book}.
Moreover, it is well known that the dipolar stray fields drive the formation of magnetic domains.\cite{1975_Battison_JPC, 1988_Pommier_JdP, 1989_Meyer_EOMOMatAppl, 2014_Karci_RSI}
In order to identify tractable evidence of magnetic domains in the quantum regime near a TF-QCP we used a symmetry-breaking field to disentangle spontaneous symmetry breaking and domain formation.

%%%%%%%%%%%%%%%%%%%%%%%%%%%%%%%%%%
\section{Magnetic phase diagram of LiHoF$_4$}

The most extensively studied material with a ferromagnetic TF-QCP is {\lhf}. Measurements of the ac susceptibility, the magnetization, and inelastic neutron-scattering established dipolar Ising ferromagnetism with a TF-QCP at $B_{\rm c}=5.1\,{\rm T}$.\cite{1996_Bitko_PRL,2005_Ronnow_Science,2007_Ronnow_PRB}
In its ordered state, {\lhf} forms simple domain patterns\cite{1970_Kaczer_IEEE, 1975_Battison_JPC, 1988_Pommier_JdP, 1989_Meyer_EOMOMatAppl, 2014_Karci_RSI, MAJorba}, with branching confined to the surface of bulk samples \cite{1988_Pommier_JdP, 1989_Meyer_EOMOMatAppl, Gabay84} and domain walls without roughening\cite{Mias05}.
Illustrated in Figs.\,\ref{fig:3}\,\textbf{a} through \ref{fig:3}\,\textbf{d} is the well-established microscopic scenario of the TF-QCP in {\lhf}, which, however, ignores magnetic domains. Easy-axis ferromagnetic order is suppressed for perfectly transverse field. Under a tilted field ($\phi\neq0$) a crossover emanates from the TF-QCP at $B^*$, because the longitudinal field-component breaks the Ising symmetry between up and down states, hence preempting spontaneous symmetry breaking.

Shown in Figs.\,\ref{fig:3}\,\textbf{e} through \ref{fig:3}\,\textbf{h} are the main results of our study including magnetic domains.
For $\phi=0$ the microscopic TF-QCP coincides with the transition from the multi-domain to the single-domain state.
However, under tilted fields  ($\phi\neq0$) the multi-domain state survives under increasing field up to a \emph{sharp second-order phase transition} at which the minority domains vanish continuously, even though the underlying properties on atomic scales undergo a crossover only.
Hence, a line of quantum critical transitions (QCTs) that are purely due to magnetic domains  (cf mesoscale QCTs) emanates from a microscopic TF-QCP.
Under increasing $\vert \phi\vert$, the transition field, $B_c$, decreases rapidly, because the field component along the easy axis favors a uniform magnetization that destabilizes the multi-domain configuration.
The qualitative difference between the purely microscopic scenario and that accounting additionally for magnetic domains, as illustrated in Figs.\,\ref{fig:3}\,\textbf{d} and \ref{fig:3}\,\textbf{h}, respectively, is corroborated by the quantitative agreement between experiment and theory.

%%%%%%%%%%%%%%%%%%%%%%%%%%%%%%%%%%
\section{Microscopic setting}

Summarized in Fig.\,\ref{fig:1} are key characteristics of the dipolar Ising ferromagnetism in {\lhf}. In the tetragonal crystal structure [Fig.\,\ref{fig:1}\,\textbf{a}, lattice constants $a=b=5.175\,{\rm \AA}$, $c=10.75\,{\rm \AA}$, space group $C^{6}_{4h}$ ($I4_1/a$)], the crystal electric field (CEF) at the location of the Ho$^{3+}$ ions exhibits an $S_4$ symmetry that originates in the surrounding F$^-$ ions. The CEF lifts the 17-fold degeneracy of the $^5I_8$ configuration, such that the ground state is a non-Kramers doublet. This doublet displays a magnetic moment along the crystallographic $c$ axis only, representing the easy magnetic axis, without appreciable magnetic anisotropy in the $ab$ plane. In our study the magnetic field was applied in the ac-plane as denoted by the angle $\phi$, Fig.\,\ref{fig:1}\,\textbf{b}.
The magnetic coupling between the Ho$^{3+}$ ions consists of long-range dipole-dipole interactions and weak exchange contributions, resulting in ferromagnetism below $T_{\rm C} = 1.53$\,K.

As a function of transverse field ($\phi=0$), the lowest CEF doublet splits with a quadratic as opposed to linear field dependence due to its non-Kramers nature, Fig.\,\ref{fig:1}\,\textbf{c} and \ref{fig:1}\,\textbf{d}. For $\phi=0$ this non-linear splitting competes with the dipolar interactions, suppressing ferromagnetism at \bc.
In addition, {\lhf} exhibits also a sizeable hyperfine coupling to the nuclear spins of the Ho ions that enhances the magnetic order at low temperatures and causes a point of inflection in the field--temperature phase boundary \cite{1996_Bitko_PRL}.
In turn, the coupled electronic and nuclear modes soften at the TF-QCP \cite{1996_Bitko_PRL, 2005_Ronnow_Science, 2007_Ronnow_PRB, 2021_Eisenlohr_PRB, 2021_Libersky_arxiv}.
For the interplay of the microscopic TF-QCP with magnetic domains, the non-Kramers nature of the Ho$^{3+}$ ions in combination with the hyperfine coupling cause a strong angle dependence providing a very sensitive test for theory, as we argue below.

%%%%%%%%%%%%%%%%%%%%%%%%%%%%%%%%%%
\section{Experimental Results}

We inferred the phase boundaries from the real and imaginary parts of the ac susceptibility, $\chi'$ and $\chi''$, as measured along the easy axis (Methods and Extended Data Fig.\,\ref{fig:EDI1}). The susceptibility for various angles $\phi$ of the static field $B$ relative to the hard axis are shown in Figs.\,\ref{fig:2}\,\textbf{a},\textbf{b} and \ref{fig:2}\,\textbf{c},\textbf{d} for $T=1.2$\,K and $62$\,mK, respectively. Further data are presented in Extended Data Fig.\,\ref{fig:EDI2} and Extended Data Table\,\ref{EDI:Theory:tab1}. The definition of $B_c$ is shown in Extended Data Fig.\,\ref{fig:EDI8}.
For what follows, we note that $\chi$ in the ordered state is dominated by the presence of ferromagnetic domains \cite{1975_Cooke_JPC, 1978_Beauvillain_PRB, 2020_Twengstrom_PRB}, while it is dominated by the microscopic response in the disordered regime.

Starting at $1.2\,{\rm K}$ and $\phi=0$, i.e., perfectly transverse field, $\chi'$ is essentially constant up to $B_c(T\!=\!1.2\,{\rm K})=3.2$\,T and decreases in the paramagnetic state, Fig.\,\ref{fig:2}\,\textbf{a}. Analyzing the field dependence of $\chi'$ for $B>B_c$ according to literature we find excellent agreement with the mean-field critical exponent $\gamma=1.02\pm0.051$ reported before \cite{1996_Bitko_PRL} (Extended Data Fig.\,\ref{fig:EDI3}).
Under tilted fields, $\phi\neq0$, we observe a well-developed phase transition, which contrasts the crossover predicted. Namely, $\chi'$ remains constant up to a sharp drop at a transition field $B_c$. In addition $\chi'$ no longer follows a power-law dependence above $B_c$ (not shown) and develops a broad maximum for large $\phi$ signalling a crossover in the disordered state. With increasing $\phi$ the transition field $B_c(\phi)$ decreases rapidly. The emergence of a sharp drop of $\chi'$ across $B_c$ for $\phi\neq0$ strongly suggests that the nature of the transition changes immediately upon tilting the field, providing evidence of the different dominant contributions to $\chi'$ in the ordered and disordered states.
Further, for fields up to $B_c$ and small $\phi$, the imaginary part of the susceptibility, $\chi''$, is finite, with a high value at low fields and a faint maximum at $B_c$. The dissipative processes underlying $\chi''$ are attributed to the motion of domain walls in the $ab$-plane in response to the small ac field along the easy axis\cite{1996_Bitko_PRL, 1975_Cooke_JPC}.
Finally, as the field direction approaches the easy axis for $\phi\to 90^{\circ}$ (Fig.\,\ref{fig:2}\,\textbf{a}), $\chi'$ is characteristic of a ferromagnet probed along its easy axis. In this limit the abrupt drop of $\chi'$ corresponds to the coercive field. Consistently, $\chi''$ at small fields increases with increasing $\phi$ and is very large up to the coercive field for $\phi\to 90^{\circ}$.

For $\phi=0$ the transition field increases for decreasing temperature and limits to $B_c=5.1$\,T in excellent agreement with the literature\cite{1996_Bitko_PRL} (Fig.\,\ref{fig:2}\,\textbf{c} for $62$\,mK, Fig.\,\ref{fig:2}\,\textbf{e} to \ref{fig:2}\,\textbf{i}, Fig.\,\ref{fig:4}\,\textbf{e}, Extended Data Fig.\,\ref{fig:EDI7}\,\textbf{a}). While the field dependence across $B_c$ at $62$\,mK and $\phi=0$ is qualitatively unchanged and characteristic of a continuous mean-field transition, $\chi'$ becomes vanishingly small below $B_{\rm fr} \sim 2$\,T. Associated with the drop in $\chi'$ is a pronounced maximum in $\chi"$, characteristic of an increase of dissipation. The behaviour of $\chi'$ and $\chi"$ around $B_{\rm fr}$ can be attributed to a freezing of the domain-wall motion at low temperatures and low fields (Extended Data Fig.~\ref{fig:EDI3}).

As shown in Fig.\,\ref{fig:2}\,\textbf{d} both $\chi'$ and $\chi''$ at $62$\,mK display the same qualitative evolution across the transition at $B_c$ as observed at higher temperatures under angles $\phi$ up to $\pm15^{\circ}$. Since $B_c$ is smaller than $B_{\rm fr}$ for $\phi \gtrsim15^{\circ}$, the susceptibility essentially vanishes suggesting that the domain walls no longer move. In turn, it is not possible to track the transition for $\phi$ exceeding $\pm15^{\circ}$.

The magnetic phase diagrams up to $\phi=15^{\circ}$ are shown in Figs.\,\ref{fig:2}\,\textbf{e} to \ref{fig:2}\,\textbf{i} (Extended Data Fig.\,\ref{fig:EDI7}\,\textbf{a}). The color shading depicting $\chi'$ underscores the location of $B_c$ and the regime of the frozen response at low fields.
With increasing $\phi$ the enhancement of $B_c$ below $\sim 0.8\,{\rm K}$ attributed to the hyperfine coupling for $\phi=0$\cite{1996_Bitko_PRL} disappears, such that the phase boundary at $\phi=10^{\circ}$ is flat as $T\to 0$.
In addition, an expanded view of $B_c$ at small fields, shown in Extended Data Fig.\,\ref{fig:EDI7}\,\textbf{b}, displays a small regime suggestive of a reentrant magnetic field dependence in the vicinity of $T_c(B=0)$. This might be consistent with fluctuations near {\tc} beyond mean-field behaviour that are quenched under an applied magnetic field.
In the low-temperature limit, $B_c$ decreases rapidly as a function of $\phi$, Fig.\,\ref{fig:2}\,\textbf{j}. To access the maximum of $B_c$ requires an alignment better $\pm0.1^{\circ}$. This sensitivity, which highlights the need to study spherical samples to ensure homogeneous demagnetizing fields, is corroborated by the jump of the susceptibility, $\Delta \chi$, at $B_c$ , Fig.\,\ref{fig:2}\,\textbf{k}.

%%%%%%%%%%%%%%%%%%%%%%%%%%%%%%%%%%
\section{Theoretical modelling}

Our main findings under tilted fields to be explained theoretically are: First, the onset of magnetic order displays the characteristics of a well-defined phase transition in $\chi'$ and $\chi''$, in stark contrast to the crossover expected in a purely microscopic scenario. Second, the transition immediately changes its character for $\phi\neq0$ and exhibits a jump in $\chi'$ with a broad maximum above $B_c$ and dissipation below $B_c$ seen in $\chi''$. Third, $B_c$ decreases rapidly with increasing $\phi$. Fourth, the temperature dependence of $B_c$ changes qualitatively when $\phi$ exceeds a few degrees, where the point of inflection and increase of $B_c$ for $T\to 0$ disappears.

The sharp phase transition we observe for $\phi\neq0$ suggests spontaneous symmetry breaking. Because the spin symmetry of the microscopic Hamiltonian for finite field is already fully broken\cite{2007_Ronnow_PRB} (Extended Data Fig.\,\ref{fig:EDI10}), an obvious candidate is translation symmetry breaking due to domain formation\cite{1970_Kaczer_IEEE, 1975_Battison_JPC, 1988_Pommier_JdP, 1989_Meyer_EOMOMatAppl, 2014_Karci_RSI, MAJorba}. This hypothesis is strongly supported by the smooth angle dependence and the qualitative similarity of $\chi$ up to $\phi=90^\circ$ (Fig.\,\ref{fig:2}\,\textbf{a}).
Going beyond previous modelling, we combine a microscopic description in tilted fields with a domain treatment, explicitly keeping domain volumes as variational parameters (Supplementary Notes\,\ref{sec:theory} and \ref{sec:landau}).
The non-Kramers ground state implies that the complete CEF scheme must be retained for the correct quantitative account\cite{1996_Bitko_PRL, 2005_Ronnow_Science}, because both the magnetization and the applied field for $\phi\neq0$ feature components along the easy \textit{and} the hard axes. A description of the Ho$^{3+}$ moment in terms of an effective spin $\nicefrac{1}{2}$ \cite{2004_Chakraborty_PRB, 2007_Ronnow_PRB, 2008_Tabei_PRB} proves to be insufficient.

For our analysis, we developed a combined and consistent account of the microscopic interactions, the mesoscopic domain energy terms, and the backaction of the stray-field on the microscopic expectation values. We start from a microscopic model of Ho$^{3+}$ ions taking into account all CEF levels, the hyperfine coupling to the Ho nuclear spin, and an inter-site interaction (Methods and Supplementary Information Note \ref{sec:microscopic}). Due to its long-range character, the dipolar coupling may be approximated by a Heisenberg interaction and decoupled in a mean-field fashion, while all local terms are treated exactly (Supplementary Note \ref{sec:meanfield}).

A large body of experimental studies in bulk samples of {\lhf} and related 3D Ising ferromagnets showed that the domains are well-described by an alternating up- down-pattern without roughening in the ordered state\cite{1935_Landau_PZS, 1970_Kaczer_IEEE} or at the critical point\cite{1986_Barker_JPC, 1986_Gabay_PRB}. To take into account the domain structure, we assumed an alternating pattern of up and down domains, with their volume ratio, $v$, as a variational parameter (Supplementary Notes \ref{sec:domains} and \ref{sec:interactions}). 

Assuming a homogeneous magnetization density in each domain, we combined the microscopic interactions and the stray fields into a consistent mean-field description of the multi-domain ferromagnet (Supplementary Note\,\ref{sec:combined}). To account for the back-action by the domains, we expressed the sum of the stray-field and the domain-wall energies as a bilinear product of expectation values of magnetic moments, $\vec{\bar J}_{1,2}$, in the domain types $1,2$,
%\begin{equation}
$
\label{eq:edom:main}
E_{\rm dom} = M \sum_{\alpha} \left( c^\alpha_1 \bar J^\alpha_1 \bar J^\alpha_1 + c^\alpha_2 \bar J^\alpha_2 \bar J^\alpha_2 + c^\alpha_{12} \bar J^\alpha_1 \bar J^\alpha_2 \right) ,
$
%\end{equation}
with $\alpha=x,y,z$ and $M$ the total number of moments. The last term, with $c^\alpha_{12}>0$, represents an effective antiferromagnetic interaction between domains which drives the multi-domain state in the vicinity of a TF-QCP. This may be viewed as emergent mesoscale antiferromagnetism, although the experimental domain pattern is not strictly periodic. In a first assessment we confirmed that the agreement between theory and experiment is not sensitive to the precise choice of the domain structure in the model (Extended Data Fig.~\ref{fig:EDI9}, Supplementary Note\,\ref{sec:energy}). We finally solved the model numerically (Supplementary Note\,\ref{sec:numerical}). 

%%%%%%%%%%%%%%%%%%%%%%%%%%%%%%%%%
\section{Mesoscale quantum criticality}

For $B<B_c$ and $\phi=0$ the up and down domains occupy equal volumes, $v^*=1/2$, and the transition from a spontaneous multi-domain to a single-domain state coincides with the microscopic TF-QCP, Fig.\,\ref{fig:4}\,\textbf{a}. This microscopic TF-QCP \emph{drives} the disappearance of domains above $B_c(\phi=0)$. For $\phi\neq0$, however, $v^*$ decreases and vanishes \textit{continuously} at $B_c (\phi\neq0)$ characteristic of a second-order phase transition, Fig.\,\ref{fig:4}\,\textbf{b} (Extended Data Fig.\,\ref{fig:EDI6}, Supplementary Note\,\ref{sec:num-pd}). In the magnetization parallel to the field, $J^B$, the suppression of minority domains results in a kink at $B_c$, which gets more pronounced with increasing $\phi$, Fig.\,\ref{fig:4}\,\textbf{c}. Interestingly, $\chi^{zz}$ drops \textit{discontinuously} at $B_c$ for $\phi\neq0$ despite of the second-order nature of the transition, because it is dominated by magnetic domains below $B_c$ and the microscopic response above $B_c$, as emphasized above Fig.\,\ref{fig:4}\,\textbf{d}. The calculated field dependence of $\chi^{zz}$, up to a single scaling parameter, is in excellent quantitative agreement with experiment (Fig.\,\ref{fig:4}\,\textbf{d}) including $\Delta\chi'$ (Fig.\,\ref{fig:2}\,\textbf{k}) and the broad maximum above $B_c$ for large $\phi$, where the latter represents a crossover similar to that at $B^*$ in the purely microscopic model.

Summarized in Fig.~\ref{fig:4}\,\textbf{e} is the calculated phase diagram for various $\phi$. The strong angle-dependence of $B_c$ (Figs.\,\ref{fig:3}\,\textbf{h} and \ref{fig:2}\,\textbf{k}) originates in the combination of the non-Kramers ground state and the hyperfine coupling (Extended Data Fig.\,\ref{fig:EDI5}, Supplementary Note\,\ref{sec:num-kram}). For $\phi=0$ the magnitude of the moment, $|J|$, is large and parallel to the $z$ axis for $B=0$, but displays a sharp cusp-like minimum at $B_c$ because the field-induced $x$ component is small. Under tilted fields the minimum of $|J|$ becomes less pronounced as it retains a finite $z$ component above $B_c$. The energy gain due to the hyperfine coupling, which scales with $\vert J \vert$, consequently favors the ferromagnetic phase and thus enhances $B_c$ strongly, however, for $\phi\lesssim5^\circ$ only.
The calculated phase diagram for a conventional moment with $J=8$ without CEF term, underscores this observation. While the nature of the transition for $\phi\neq0$ remains unchanged, changes of $B_c$ as a function of $\phi$ are significantly weaker and the hyperfine coupling influences $B_c$ only marginally, because $|J|$ is now independent of the field, and the hyperfine coupling no longer favors a particular moment direction.

Taken together, we identify two immanently different though closely connected forms of field-driven quantum criticality. For $\phi=0$, dipolar ferromagnetism terminates in a well-understood microscopic Ising TF-QCP. Due to the long-ranged nature of the dipolar interactions this transition is mean-field with critical exponents $\alpha=0$, $\beta=1/2$, $\gamma=1$, $\delta=3$. However, the divergence of the correlation length on the ordered side of the TF-QCP will be ultimately limited by the domain size.
In contrast, for $\phi\neq 0$ the transition involves magnetic domains only, where the continuous suppression of the domain ratio $v^*$ (Fig.\,\ref{fig:4}\,\textbf{b}) drives a mesoscale QCP. This transition is highly unusual as it is dominated by the long-ranged character of the stray fields. Formally, one may assign critical exponents $\alpha=0$, $\beta=1$, $\gamma=0$, $\delta=1$, where $\beta=1$ reflects the linear behavior of $v^*$ near $B_c$ and $\gamma=0$ is consistent with the jump in $\chi^{zz}$.

%%%%%%%%%%%%%%%%%%%%%%%%%%%%%%%%%%
\section{Concluding remarks}

Our study benefits from the simplicity of 3D Ising ferromagnetic order with long-range dipolar interactions, for which mean-field behaviour is expected and observed (Extended Data Fig.\,\ref{fig:EDI3} and Ref.\,\onlinecite{1996_Bitko_PRL}) with additional logarithmic corrections in the magnetization and susceptibility at the upper critical dimension.\cite{1978_Beauvillain_PRB, Griffin80} Likewise, we find the phase diagram of {\lhf} to be quantitatively consistent with mean--field theory. Yet, there are several regimes where fluctuations beyond mean-field must be anticipated, for instance, when the size of the minority domains approaches microscopic distances near $B_c$, or near the reentrant phase boundary at small $\phi$ (cf. Extended Data Fig.\,\ref{fig:EDI7}\,\textbf{b}). This motivates the development of new methods to study mesoscale texture in near quantum criticality and quantum critical mode softening of carefully tailored mesoscale textures in the future.

Without defect-related pinning the character of the mesoscale quantum phase transitions we observe under tilted fields remains qualitatively unchanged up to $\phi=\pm90^{\circ}$. They connect, hence, continuously with the coercive field along the easy axis, suggesting that pure anisotropic ferromagnets display mesoscale QCPs generically. In turn, the interplay of pinning with mesoscale QCPs offers a new perspective of quantum annealing and generic random field physics.
Deep connections exist also with ferroelectrics and multiferroics\cite{2008_Hubert_Book, 2020_Evans_PSR} which we expect to display related mesoscale quantum phase transitions. However, in ferroelectrics and multiferroics the search for microscopic quantum transitions is still at an early stage. \cite{2017_Chandra_RPP, 2020_Coak_PNAS} The observation of mesocale QCPs in {\lhf} finally motivates also to revisit the large body of unresolved inconsistencies, including instances of frozen spin order, observed at quantum phase transitions of metallic magnets.\cite{2001_Pfleiderer_Nature, 2015_Schmakat_EPJST, 2016_Brando_RMP, 2021_Rana_PRB} The identification of emergent mesoscale quantum criticality in {\lhf} thus promises the discovery of a rich variety of materials properties in the future -- some of which even without classical analogue.

% word count
% 21.5.2022 at 12:52  	2699
% 29.5.2022 at 17:52	2708

%\noindent \textbf{[requested word count: 2700; current count: 3321]}
% cut by 48 lines;

%%%%%%%%%%%%%%%%%%%%%%%%%%%%%%%%%%%

\newpage
\section*{References}
%\bibliography{LiHoF4-v2}

%%%%%%%%%%%%%%%%%%%%%%%%%%%%%%%%%%%

%%%%%%%%%%%%%%%%%%%%%%%%%%%%%%%%%%

%%%%%%%%%%%%%%%%%%%%%%%%%%%%%%%%%%
\newpage
\section*{Main figures}

%%%%%%%%%%%%%%%%%%%%%%%%%%%%%%%%%%%
\clearpage \thispagestyle{empty}

\captionsetup[figure]{labelfont={bf},name={Fig.},labelsep=space}

\begin{figure}[ht]
	\centerline{\includegraphics[width=0.4\textwidth,clip=]{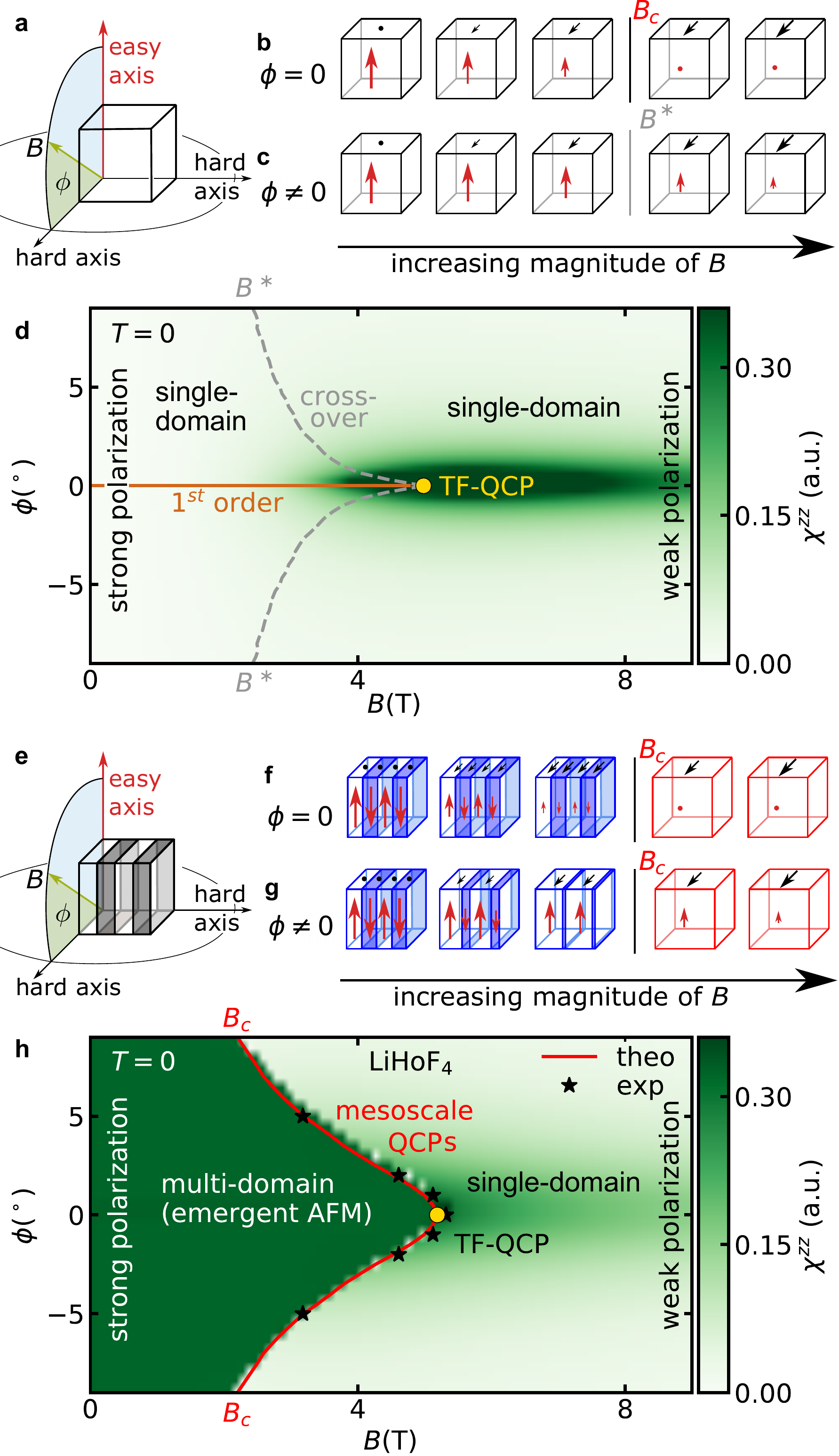}}
\linespread{1.0}\selectfont{}
\caption{\raggedright
{\bf $|$ Zero-temperature phase diagram of a microscopic transverse-field Ising model and of a domain model for {\lhf} as a function of field strength, $B$, and field orientation, $\phi$.}
Cubes serve to depict magnetisation components along the easy magnetic axis (red) and the hard magnetic axes (black) as defined in panels \textbf{a} and \textbf{e}. The field direction encloses the angle $\phi$ with the hard magnetic axis, i.e., transverse-field conditions correspond to $\phi=0$. The color shading in \textbf{d} and \textbf{h} represents the magnetic susceptibility for the easy axis. Under increasing field a strong polarization along the easy axis is suppressed. \textbf{b}, \textbf{c}, \textbf{d}, Purely microscopic scenario not accounting for stray fields and the formation of magnetic textures. At $\phi=0$ a line of first-order phase transitions terminates in the transverse-field quantum critical point (TF-QCP). For $\phi\neq0$ a crossover is observed at a characteristic field $B^*$ depicted by the grey-dashed line. \textbf{f}, \textbf{g}, \textbf{h}, Identical microscopic model as considered in panels \textbf{a} through \textbf{d}, however, accounting for the effects of stray magnetic fields and the concomitant formation of a multi-domain state in terms of an emergent antiferromagnetism (AFM). A well-defined phase transition separates the multi-domain state (emergent  AFM) at low field from the single-domain (weakly polarized) state at high fields. The ratio of the domain volumes results in a continuous (second order) suppression of the multi-domain state for $\phi\neq0$ forming a line of mesoscale quantum critical points (QCPs). This transition coincides with the microscopic TF-QCP at $\phi=0$. Excellent qualitative and quantitative agreement with experiment is observed, where the data points shown here were inferred from a linear extrapolation to zero temperature.
%\newline
%label: fig:3; word count: 232
}
\label{fig:3}
\end{figure}

%%%%%%%%%%%%%%%%%%%%%%%%%%%%%%%%%%%
\clearpage \thispagestyle{empty}

\begin{figure*}[ht]
	\centerline{\includegraphics[width=0.7\textwidth,clip=]{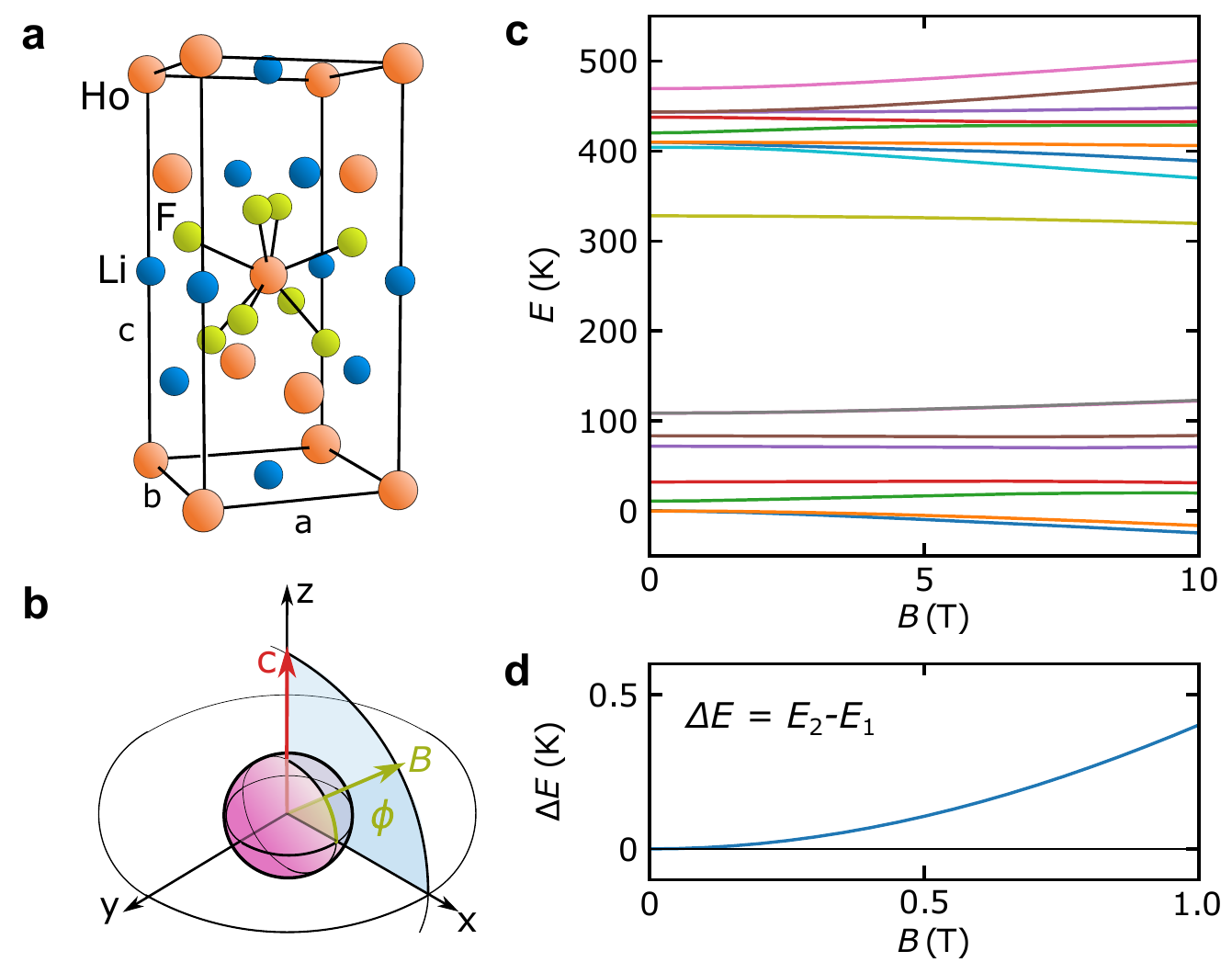}}
\linespread{1.0}\selectfont{}
\caption{\raggedright
{\bf $|$ Crystal structure of {\lhf} and associated crystal electric field (CEF) scheme under transverse magnetic field.}
\textbf{a}, Crystallographic unit cell of {\lhf}. The $c$-axis represents the easy magnetic axis; magnetic anisotropies in the $ab$-plane are vanishingly small. Data were recorded in the $ac$-plane.
\textbf{b}, Definition of the angle $\phi$ relating the field direction to the crystallographic orientation. In our experimental studies a spherical sample was used to ensure uniform demagnetizing fields across the sample volume.
\textbf{c}, CEF scheme of single ion Ho$^{3+}$ in the crystalline environment of {\lhf} under transverse field.
\textbf{d}, Splitting of lowest-lying doublet under transverse magnetic field, exhibiting a quadratic, i.e., nonlinear field dependence as a key characteristic of the non-Kramers ground state. To account for our experimental data the entire set of CEF levels must be taken into account, going well beyond an effective spin 1/2 description \cite{1996_Bitko_PRL, 2004_Chakraborty_PRB, 2005_Ronnow_Science, 2007_Ronnow_PRB, 2008_Tabei_PRB}. In addition, the formation of magnetic domains due to stray-fields proves to be essential.
%\newline
%label: fig:1
}
\label{fig:1}
\end{figure*}

%%%%%%%%%%%%%%%%%%%%%%%%%%%%%%%%%%%
\clearpage \thispagestyle{empty}

\begin{figure*}
	\centering\includegraphics[width=0.55\textwidth,clip=]{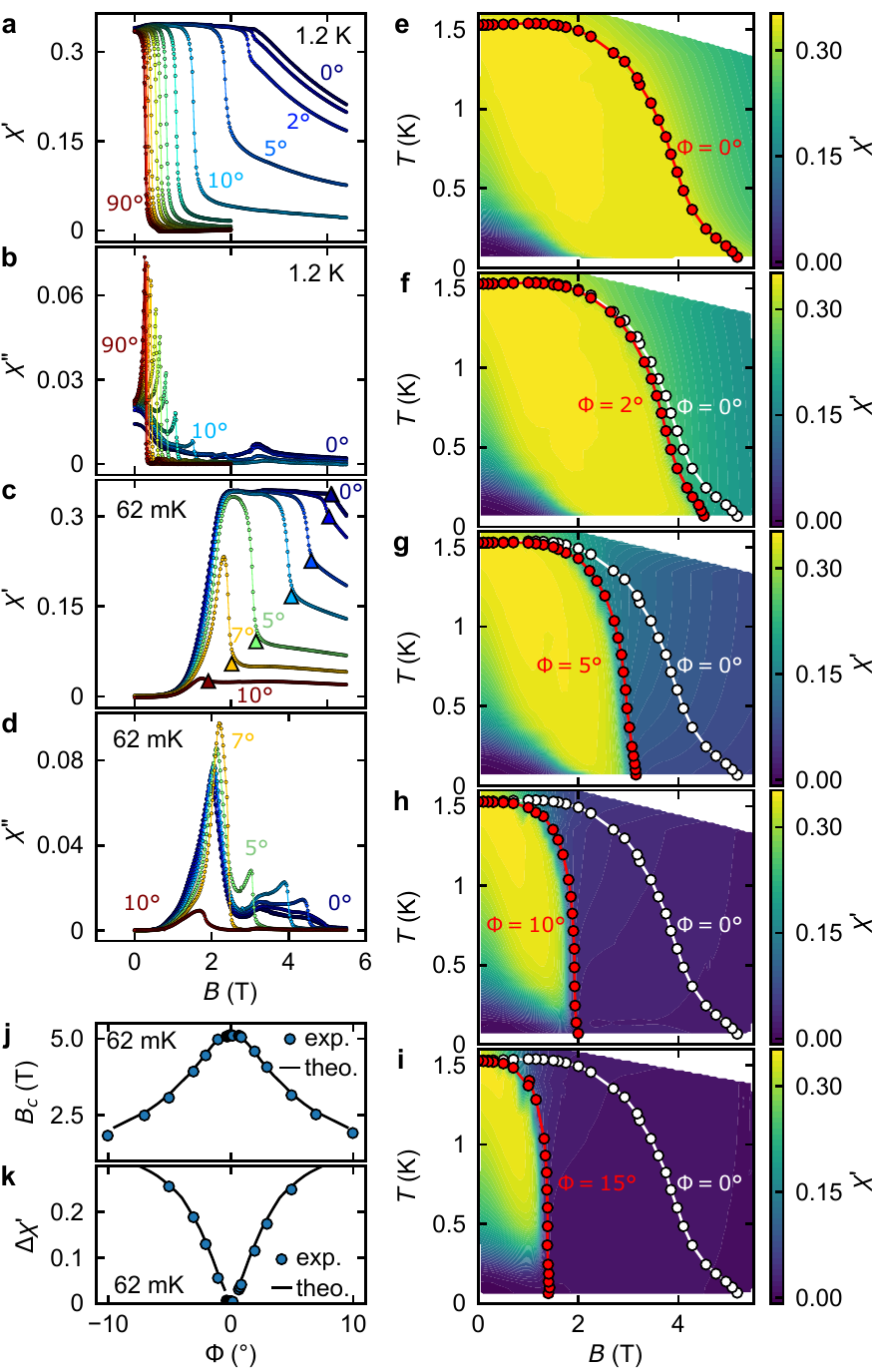}
\linespread{1.0}\selectfont{}
\caption{\raggedright
{\bf $|$ Experimentally recorded susceptibility and temperature versus field phase diagram of {\lhf} at various field orientations as denoted by the tilt angle $\phi$.}
\textbf{a}, \textbf{b}, Real and imaginary parts of the ac susceptibility, $\chi'$ and $\chi''$, along the easy axis of {\lhf} at 1.2\,K as a function of a static magnetic field $B$ under different angles between $\phi=0$ and $\phi=90^{\circ}$, i.e., transverse and parallel to the easy axis. The color shading of the data serves to enhance contrast.
\textbf{c}, \textbf{d}, $\chi'$ and $\chi''$ at 62\,mK as a function of $B$ under different angles between $\phi=0$ and $\phi=15^{\circ}$. Triangles mark the transition as defined in Extended Data Fig.\,\ref{fig:EDI8}; the small but finite width of the transition may be attributed to surface pinning and other minor experimental artefacts.
\textbf{e} through \textbf{i}, Temperature versus magnetic field phase diagrams for different field orientations $\phi$. The color shading denotes the magnitude of the real part of the susceptibility, $\chi'$, along the easy axis. Panels \textbf{f} through \textbf{i} display for comparison the phase transition line for $\phi=0$.
\textbf{j}, Critical field at 62\,mK as a function of tilting angle $\phi$ with respect to the ideal transverse field orientation ($\phi=0$). The line is based on the theoretical model described in the text.
\textbf{k}, Change of the real part of the susceptibility, $\Delta \chi'$, as a function of tilting angle $\phi$, where $\Delta \chi'$ is defined as shown in Extended Data Fig.\,\ref{fig:EDI8}. Data are normalized with respect to the transverse field orientation ($\phi=0$). Data were recorded at 62\,mK. The line is based on the theoretical model described in the text, where the calculated susceptibility curves were scaled by the same constant such that they matched the experimental value of the plateau in the multidomain state (cf. Fig.\,\ref{fig:4}\,\textbf{d}).
%
%\newline
%label: fig:2
}
\label{fig:2}
\end{figure*}

%%%%%%%%%%%%%%%%%%%%%%%%%%%%%%%%%%%
\clearpage \thispagestyle{empty}

\begin{figure}[ht]
	\centerline{\includegraphics[width=0.65\textwidth,clip=]{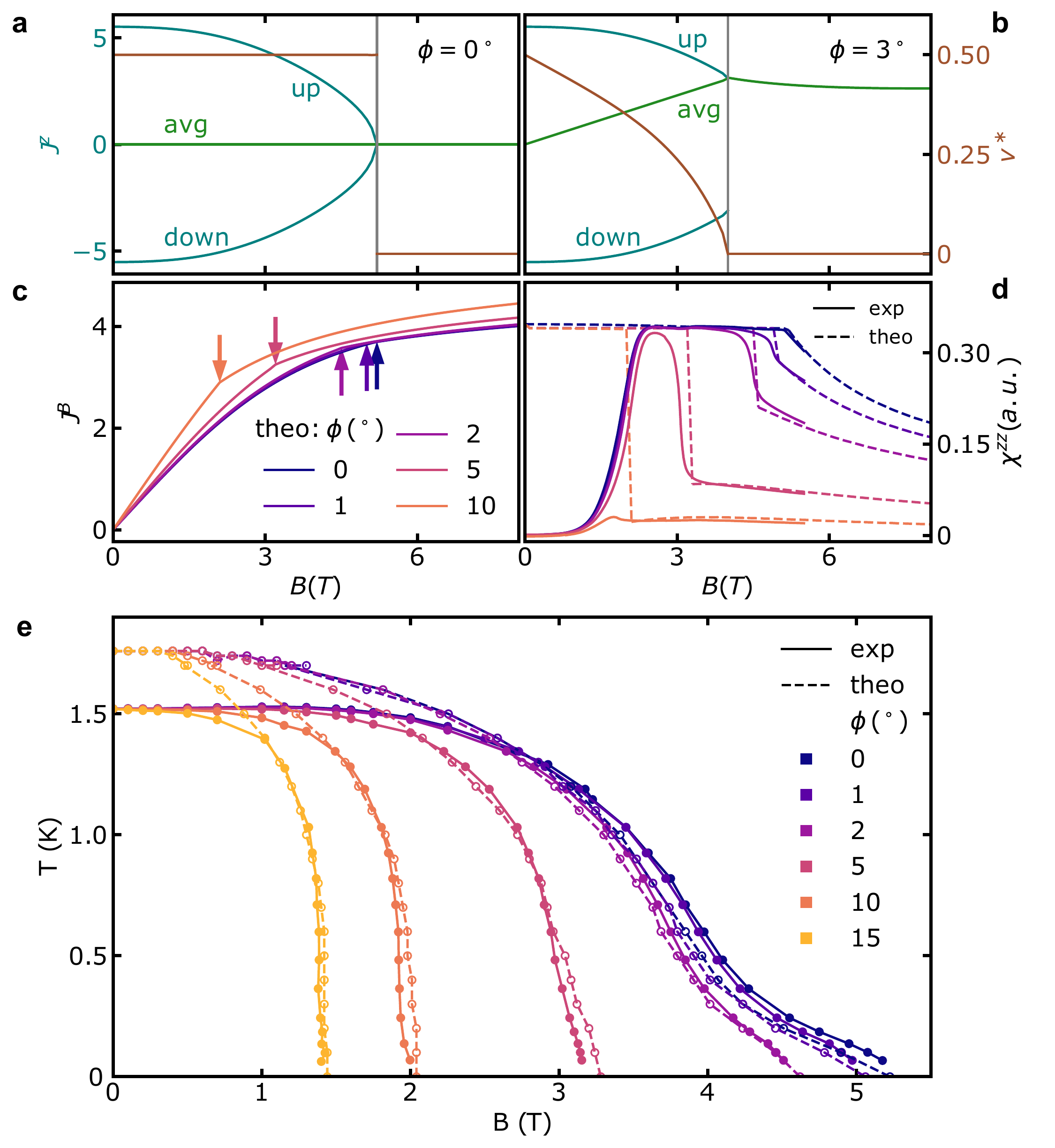}}
\linespread{1.0}\selectfont{}
\caption{\raggedright
{\bf $|$ Key aspects of the multi-domain state (emergent mesoscale AFM) in the vicinity of the microscopic transverse-field quantum critical point (TF-QCP) in {\lhf}.}
\textbf{a}, \textbf{b}, Calculated easy-axis magnetic moment, $J_z$, for the different domains (up/down) and the domain-average (avg) under field orientations $\phi=0$ and $\phi=3^\circ$. Also shown is the domain ratio, $v^*$.
For $\phi=0$ the domain types assume equal ratio, $v^*=1/2$, up to the transition at $B_c$. In contrast, for $\phi\neq0$ the domain ratio $v^*$ differs for finite fields and approaches zero at $B_c$ continuously, i.e., the minority domains shrink monotonically and vanish at $B_c$ in a second-order transition.
\textbf{c}, Calculated magnetization of {\lhf} along the field direction, $J_B$, and for different tilt angles $\phi$. For $\phi=0$ this corresponds to the magnetization along the hard axis. For $\phi\neq0$ the magnetization $J_B(B)$ displays a kink at $B_c$. As the tilt angle decreases the change of slope at the kink vanishes.
\textbf{d}, Calculated real-part of the easy-axis susceptibility, $\chi'^{zz}$, and corresponding experimental data. Note that $\chi'^{zz}$ is dominated by the multi-domain state below $B_c$ and the microscopic single-domain response above $B_c$. The discontinuous drop in $\chi'^{zz}$ is hence consistent with a continuous transition at $B_c$. It coincides with the kink in $J_B$ at $B_c$ along the hard axis. All of the calculated susceptibility curves were scaled by the same constant such that they matched the experimental value of the plateau in the multidomain state.
\textbf{e}, Quantitative comparison of the calculated and the experimental temperature versus magnetic field phase diagram for different tilt angles $\phi$. Squares in the legend serve to denote the color shading used for experiment and theory. The actual data points are denoted by circles. Excellent qualitative and quantitative agreement is observed below $\sim1.2\,{\rm K}$ for all $\phi$. The calculated Curie temperature is $\sim 20\,\%$ larger than experiment (cf. Extended Data Fig.\,\ref{fig:EDI7}.) We attribute the difference to fluctuation effects beyond the mean-field model used in our analysis.
%\newline
%label: fig:4
}
\label{fig:4}
\end{figure}

%%%%%%%%%%%%%%%%%%%%%%%%%%%%%%%%%%
\clearpage
\newpage
\section*{Methods}

%References \cite{1963_Gemperle_pss, 1963_Gemperle_pss, 1990_Koetzler_PRL}

%%%%%%%%%%%%%%%%%%%%%%%%%%%%%%%%%%
\textbf{Considerations on the methodology}

The scope of our study concerned the presence and relevance of mesoscale textures in a regime of strong quantum fluctuations. For a combined experimental and theoretical investigation, this question was narrowed down on magnetic materials connecting the fields of quantum phase transitions and textures in magnetic materials. 

A cornerstone of the field of quantum phase transitions in magnetic materials represents the development of a continuum quantum field theory using a Landau-Ginzburg-Wilson functional, taking into account quantum fluctuations of the order parameter. As established by the seminal work of Hertz \cite{1976_Hertz_PRB} and Millis \cite{1993_Millis_PRB}, in this approach the time direction may be interpreted as an additional space direction, such that the quantum theory of the zero temperature phase transition in $d$ dimensions is equivalent to a classical theory in $d_{\rm eff}=d+z$ dimensions, where $z$ is the dynamical exponent. In turn, mean-field critical behaviour is expected when the effective dimension is at or above the well-known upper critical dimension $d_{\rm c}=4$. These mean-field predictions of quantum critical transitions triggered a large body of developments.\cite{2011_Sachdev_Book, 2003_Vojta_RPP, 2007_HvL_RMP, 2009_Pfleiderer_RMP, 2016_Brando_RMP, 2018_Vojta_RPP} Theoretical work in the preceding decades that took into account material-specific properties revealed corrections beyond mean field behaviour. For instance, in itinerant-electron magnets the additional presence of fermionic degrees of freedom was found to cause additional soft-modes \cite{Belitz_PRL_2002}. Another example are Berry phase contributions, which were found to be important at quantum phase transitions of frustrated spin systems.

In magnetic materials the search for mesoscale textures may allude to magnetic domains, on the one hand, and strictly periodic, long-wavelength modulations, on the other hand. As the latter effectively represent conventional antiferromagnets, the focus of our study was placed on magnetic domains, where two facets access potentially new directions. First, while being non-periodic on long distances they break translation symmetry. Second, they are driven by long-range interactions, notably dipolar stray fields. Technological and scientific challenges pursued in studies of such dipolar field-driven non-periodic textures comprise key characteristics of their size, shape, creation, decay, and dynamical properties including collective excitations, as well as their manipulation and motion using spin and charge currents.\cite{2008_Hubert_Book} A new development in recent years in this field concerned topologically non-trivial spin textures such as skyrmions, merons, hopfions and certain disclinations in chiral and geometrically frustrated magnets.\cite{2009_Muehlbauer_Science, 2013_Nagaosa_NN, 2020_Back_JPD} While a large body of studies has addressed such topologically non-trivial textures from a purely classical point of view, only a small number of studies addressed their quantum properties \cite{2019_Ocha_IJMPB, 2019_Lohani_PRX}. 

Given the well-known complexities of magnetic domains and their sensitivity to very weak perturbing potentials, three key aspects are important in the selection of a suitable material: (i) the shape and the size of the magnetic domains, (ii) differences of the domains in the bulk and at the surface, and (iii) the character of the domain walls. Taking these aspects into account we decided here to focus on the quantum critical point in insulating 3D Ising ferromagnets under transverse field (TF-QCP) \cite{2011_Sachdev_Book, 2013_Suzuki_Book, 2016_Dutta_Book}. As summarized in Supplementary Note \ref{sec:domains}, either sheet-like or cylinder-like domain patterns exist in the bulk that are energetically known to be almost equivalent.\cite{1963_Gemperle_pss,1970_Kaczer_IEEE} Moreover, branching is confined to the surface of bulk samples, such that the contributions in bulk samples are small.\cite{Gabay84, Gabay85, 1986_Barker_JPC} Finally, the domain walls do not exhibit roughening.\cite{1986_Gabay_PRB, 1990_Koetzler_PRL, Mias05} This way the rather simple modelling of the effects of domains in a bulk sample is justified.

As domain formation is driven by dipolar stray fields, it is of utmost importance to perform experimental studies in a setting where these stray field are as homogeneous as possible. We decided therefore to investigate spherical samples (measurements on cuboid samples exhibited massive smearing effects such that the data are not useful\cite{2010_Legl_PhD}). While microscopic studies of the domains inside the bulk of samples are desirable, the necessary experimental methods are still at an infant stage. For instance, neutron depolarization radiography in combination with grating interferometry might allow to obtain such information in the future.\cite{2009_Kardjilov_NatPhys, 2010_Pfleiderer_JLTP, 2021_Jorba_PhD} However, substantial improvements of spatial resolution would be necessary and experimental implementations that permit studies at milli-Kelvin temperatures under changing field direction, representing major challenges for studies using polarized neutrons. Another option may be x-ray imaging of magnetic structures\cite{2017_Donnelly_Nature}, where the necessary analysis algorithms would have to be developed as well as experimental set-ups that permit studies at at milli-Kelvin temperatures under changing field direction. Finally, use of imaging methods at sample surfaces to gather information on the bulk properties would require the development new experimental strategies to avoid surface-induced modifications of domain structures such as branching.

Meeting these considerations our study focussed on the relevance of comparatively simple domains in the vicinity of a well understood microscopic quantum critical transition to obtain a tractable point of reference for future studies of materials with more complex domains. Such future directions on magnetic domains in the quantum regime are well beyond the scope of our study. They comprise, for instance: (i) the role of different magnetic anisotropies, (ii) the effects of spin-orbit coupling, (iii) the effects of itinerant electrons, (iv) the effects of frustration, or (v) the role of sample surface and sample shape.

Mesoscale textures at quantum phase transitions may become gateways for new directions of condensed matter research in the future as follows:
\begin{compactitem}
\item Exploration of quantum phenomena at the transverse field ferromagnetic quantum critical points beyond the mean-field character we observed in {\lhf}, as the objective of our study concerned the mere identification of evidence that such mesoscale quantum phase transitions exist at all. Closely related is the putative relevance of such textures in the regime of quantum spin glasses and for quantum annealing.\cite{2011_Sachdev_Book, 2013_Suzuki_Book, 2016_Dutta_Book, 2019_Silevitch_NatComms}
\item Investigation of the relevance of mesoscale textures in the context of unresolved issues at conventional quantum phase transitions in magnetic materials. This includes, for instance, quantum criticality in itinerant ferromagnets, where the observation of diffusive charge dynamics has been an unresolved puzzle for several decades.\cite{2001_Pfleiderer_Nature, 2007_HvL_RMP, 2010_Pfleiderer_JLTP, 2016_Brando_RMP, 2018_Vojta_RPP, 2015_Schmakat_EPJST} In these systems, it has been speculated that mesoscale textures may even represent a generic escape route for microscopic quantum criticality.\cite{2009_Conduit_PRL, 2015_Abdul_NatPhys, 2018_Friedemann_NatPhys} Similarly, there is increasing evidence for an abundance of textures and spin frozen behavior in the vicinity of quantum phase transitions.\cite{2004_Pfleiderer_Nature, 2013_Ritz_Nature, 2021_Seifert_PhD, 2021_Jorba_PhD, 2021_Rana_PRB, 2020_Green_ARCMP}
\item Search for and exploration of mesoscale quantum phase transitions in materials with topologically non-trivial spin and charge textures such as skyrmions, merons, and hopfions expanding on seminal theoretical work addressing quantum skyrmionics.\cite{2019_Ocha_IJMPB, 2019_Lohani_PRX}
\item Search for and exploration of mesoscale quantum phase transitions in materials with non-magnetic order parameters. For instance, it is well-known since the early work of de Gennes\cite{1963_deGennes_SSC} nearly seven decades ago, that ferroelectrics are also model systems par excellence of the transverse field Ising physics. Accordingly, we expect similar behavior near ferroelectric quantum phase transitions,\cite{2014_Rowley_NatPhys, 2017_Chandra_RPP, 2020_Coak_PNAS} and, on a more general note, multiferroics and related compounds.\cite{2020_Evans_PSR}
\end{compactitem}

%%%%%%%%%%%%%%%%%%%%%%%%%%%%%%%%%%
\textbf{Experimental techniques}

Several single-crystal {\lhf} samples were purchased from AcalBfi/Altechna Germany. The optical appearance, characterization by x-ray and neutron Laue diffraction, as well as magnetic properties measured in a Quantum Design physical properties measurement system consistently confirmed excellent sample quality in agreement with the literature. For our measurements single-crystalline pieces were carefully ground and polished into spheres to reduce sample-shape related inhomogeneities of the demagnetizing fields.\cite{2017_Twengstrom_PRM, 2020_Twengstrom_PRB} In our study a sample with a diameter of $d=2.8\,{\rm mm}$ was investigated. After grinding and polishing the quality of the samples was remeasured and found to be unchanged excellent.
The use of spherical samples here proved to be decisive. Using instead cuboid-shaped samples, we observed strongly smeared out transitions for $\phi=0$ and $\phi\neq0$ \cite{2010_Legl_PhD} rendering a tractable interpretation essentially impossible.

In our study we measured the ac susceptibility along the easy magnetic axis of {\lhf}, probing the ferromagnetic order parameter by means of a bespoke miniature susceptometer comprising a primary coil with a balanced pair of secondaries \cite{2019_Rucker_RSI} (cf. Extended Data Fig.\,\ref{fig:EDI1}). Data were recorded at an excitation frequency $f=511\,{\rm Hz}$ and an excitation amplitude $B_{\rm ac}\approx13\,\mu{\rm T}$.
The susceptibility data were calibrated by means of measurements of the longitudinal susceptibility, i.e., $\phi=90^{\circ}$, using a Quantum Design physical properties measurement system down to 2.3\,K \cite{2018_Rucker_PhD}. All susceptibility data are reported in SI units following convention.
A JT Oxford Instruments dilution refrigerator in combination with a two-axes American Magnetics vector magnet (9\,T, 4.5\,T) was used for the measurements under transverse fields down to mK temperatures. The sample temperature was tracked with calibrated RuO sensors purchased form Lakeshore.

The susceptometer was mounted such that its orientation was perpendicular to the vertical axis of the dilution refrigerator.
For the measurements of the susceptibility the sample was attached to the sapphire rod with GE varnish such that the easy magnetic axis was parallel to the susceptometer coils and thus perpendicular to the axis of the dilution refrigerator. The orientation of the sample with respect to the sapphire rod and  susceptometer coils was confirmed to be better than a degree using x-ray Laue diffraction.

The precise orientation of the sample with respect to the susceptometer coils and the vector magnet was determined and adjusted by means of a two stage procedure. At first the critical field was mapped out at $T  = 62 \,{\rm mK}$ as a function of the magnets field angle for an orientation of the dilution refrigerator such that the easy axis was roughly perpendicular to the plane of the vector magnet. Following this the dilution refrigerator with the susceptometer attached was rotated by 90$^{\circ}$ with respect to the vertical axis and thus the plane of the two-axes magnet. This effectively brought the easy axis close to the plane of the two-axes magnet. Next, the critical field was mapped out again as a function of the field orientation, analogous to Fig.\,\ref{fig:2}. Based on the data recorded for the two magnetic field planes, the precise orientation of the dilution refrigerator with respect to the plane of the two-axes magnet was determined and the vertical axis adjusted such that the easy axis was accurately located in the plane of the two-axis magnet.

Small tilt angles of less than a few degrees of the easy axis with respect to the vertical axis of the dilution refrigerator were finally inferred from measurements through a wide range of angles. For the sake of clarity only the angle $\phi$ of the field orientation with respect to the easy crystallographic axis is reported in our manuscript as depicted in Fig.\,\ref{fig:1}\,\textbf{b}.

Based on a careful estimate of the combination of systematic uncertainties in the alignment procedure and the statistical error of the vector field, a conservative estimate of the accuracy of the angle $\phi$ of the field orientation is $\Delta\phi=\pm0.02^{\circ}$. The uncertainties of the magnitude of the applied magnetic field corresponded to the jitter of the power supply, well below the detection limit. The error of the temperature stability corresponded to better $\pm0.1\%$, whereas the absolute accuracy of the calibration of the thermometers corresponded to the values provided by the commercial supplier.

%%%%%%%%%%%%%%%%%%%%%%%%%%%%%%%%%%
%\newline
\textbf{Theoretical modelling}

Our theoretical model starts from a microscopic description of the full local Hilbert space of electronic and nuclear moments and a mean-field treatment of the interactions between the electronic moments, as in earlier work.\cite{1996_Bitko_PRL,2005_Ronnow_Science,2005_Schechter_PhysRevLetta} This is supplemented by the magnetostatics of a mesoscopic periodic arrangement of domains of variable size and magnetization. Importantly, the combined theory thus contains the interplay between microscopic and mesoscopic degrees of freedom. Full details of the theoretical framework beyond the summary presented in the main text and here may be found in Supplementary Information Notes\,\ref{sec:landau} through \ref{sec:numerical}.

Microscopically LiHoF$_4$ is described by a transverse-field exchange model augmented by crystal electric fields $\vcf$ and hyperfine coupling $A$ to nuclear spins,
\begin{eqnarray}
H_{\rm mic}
&=&  - K \sum_{\langle ij\rangle} \vec J_i \cdot \vec J_j + \sum_i \big[\vcf(\vec J_i) + A \vec J_i \cdot \vec I_i \big] \nonumber \\
&-& \mu_B \vec B \cdot \sum_i (g \vec J_i + g_N \vec I_i),
\end{eqnarray}
where $g$ ($g_N$) are the electronic (nuclear) Land\'{e} factors, and the external field $\vec B = B (\sin\phi,0,\cos\phi)$ is varied in its strength $B$ and tilt angle $\phi$. The electronic moments interact via dipole-dipole and exchange interactions; their combined effect is captured by the nearest-neighbor ferromagnetic coupling $K$. The Ising anisotropy is exclusively contained in the single-ion crystal field $\vcf$. The electronic and nuclear moments with $J=8$ and $I=7/2$, respectively, form a $(17 \times 8)$-dimensional Hilbert space on each site, which is kept in its entirety,\cite{1996_Bitko_PRL, 2005_Ronnow_Science, 2005_Schechter_PhysRevLetta} in contrast to some previous works which used a projection to an effective $2 \times 8$ dimensional model.\cite{2004_Chakraborty_PRB, 2018_McKenzie_PRB}
In order to solve $H_{\rm mic}$, the interaction is treated in a mean-field approximation, with site-independent mean fields $\bar J^x, \bar J^z$.

Domain formation is incorporated by introducing two types of domains, 1 (up) and 2 (down), and by accounting for stray fields. The domains form an alternating pattern on mesoscopic scales and are assumed to be sheets stacked in the $y$ direction, as sketched in Extended Data Fig.\,\ref{fig:EDI6}\,\textbf{a}. Assuming a homogeneous magnetization in each domain, the sum of total stray-field and domain-wall energies can be expressed as a function of the domain magnetizations, where it takes the form of an effective antiferromagnetic interaction (Eq.\,\ref{eq:edom:main}) between domains of type 1 and 2. In the spirit of the mean-field approximation, this interaction is decoupled in a fashion similar to the microscopic spin-spin coupling. Estimates for the domain-wall energy density \cite{2009_Biltmo_EPL} and the domain sizes \cite{1988_Pommier_JdP, 1989_Meyer_EOMOMatAppl, 2014_Karci_RSI} are taken from the literature. Together, this results in a consistent mean-field treatment of the multi-domain ferromagnet, with a separate set of mean fields for each domain type. Importantly, it includes the volume ratio $v$ of the domain types 1 and 2, $v=D_2/(D_1+D_2)$, as variational parameter. The single-domain solution is obtained in the limit of either identical mean fields in domains of type 1 and 2 or vanishing volume of one domain type.

%%%%%%%%%%%%%%%%%%%%%%%%%%%%%%%%%%
%%%%%%%%%%%%%%%%%%%%%%%%%%%%%%%%%%
\newpage
\section*{References}
%\bibliography{LiHoF4-v2}

%%%%%%%%%%%%%%%%%%%%%%%%%%%%%%%%%%

%%%%%%%%%%%%%%%%%%%%%%%%%%%%%%%%%%
\newpage
\section*{Acknowledgements}

We wish to thank P. B\"oni, C. Castelnovo, T. Enns, M. Garst, P. Jorba Cabre, M. Knap, J. Knolle, M. Lampl, S. Legl, M. Meven, R. Moessner, H. Ronnow, J. Schmalian, S. S\"aubert, F. Pollmann, and W. Zwerger for support and discussions.
We acknowledge also support by S. Mayr and the mechanical workshop at the Physik-Department of the Technical University of Munich.
Financial support from the Deutsche Forschungsgemeinschaft (DFG, German Research Foundation) through the Munich Center for Quantum Science and Technology (EXC 2111, project-id 390814868), the W\"urzburg-Dresden Cluster of Excellence on Complexity and Topology in Quantum Matter -- \textit{ct.qmat} (EXC 2147, project-id 390858490), SFB 1143 (project-id 247310070), and TRR80 (project-id 107745057) is gratefully acknowledged.
This project has received funding from the European Research Council (ERC) under the European Union's Horizon 2020 research and innovation programme (grant agreement No 788031, ExQuiSid).

%%%%%%%%%%%%%%%%%%%%%%%%%%%%%%%%%%%

\section*{Author Contributions}
A.W., F.R., C.D., M.K. and C.P. conducted the measurements.
A.W., F.R. and C.P. analyzed the data.
C.P. proposed this study.
H.E. and M.V. developed the theoretical model.
A.W., H.E., M.V, and C.P. conceived the interpretation and wrote the manuscript.
All authors discussed the data and commented on the manuscript.
%\newpage

%%%%%%%%%%%%%%%%%%%%%%%%%%%%%%%%%%%

\section*{Additional Information}

In the supplementary notes we report detailed information on the theoretical model and the theoretical analysis, comprising the Landau theory of the transverse Ising transition under tilted magnetic fields, the microscopic Hamiltonian of {\lhf} and its mean-field approximation, the account of multi-domain energetics and the numerical evaluation of our combined model connecting the microscopic and mesocale phenomena.

%%%%%%%%%%%%%%%%%%%%%%%%%%%%%%%%%%

\section*{Data availability}
Materials and additional data related to this paper are available from the corresponding authors upon reasonable request.

%%%%%%%%%%%%%%%%%%%%%%%%%%%%%%%%%%%

\section*{Competing interests}
The authors declare no competing interests.

%%%%%%%%%%%%%%%%%%%%%%%%%%%%%%%%%%%

\section*{Correspondence and requests for materials}
Correspondence and requests for materials should be addressed to C.P. or M.V.

%%%%%%%%%%%%%%%%%%%%%%%%%%%%%%%%%%
\newpage

\section*{Extended Data}

\setcounter{figure}{0}
\captionsetup[figure]{labelfont={bf},name={Extended Data Fig.},labelsep=space}

%\iffalse

%%%%%%%%%%%%%%%%%%%%%%%%%%%%%%%%%%%
\clearpage \thispagestyle{empty}

\begin{figure*}[h]
	\centerline{\includegraphics[width=0.5\textwidth,clip=]{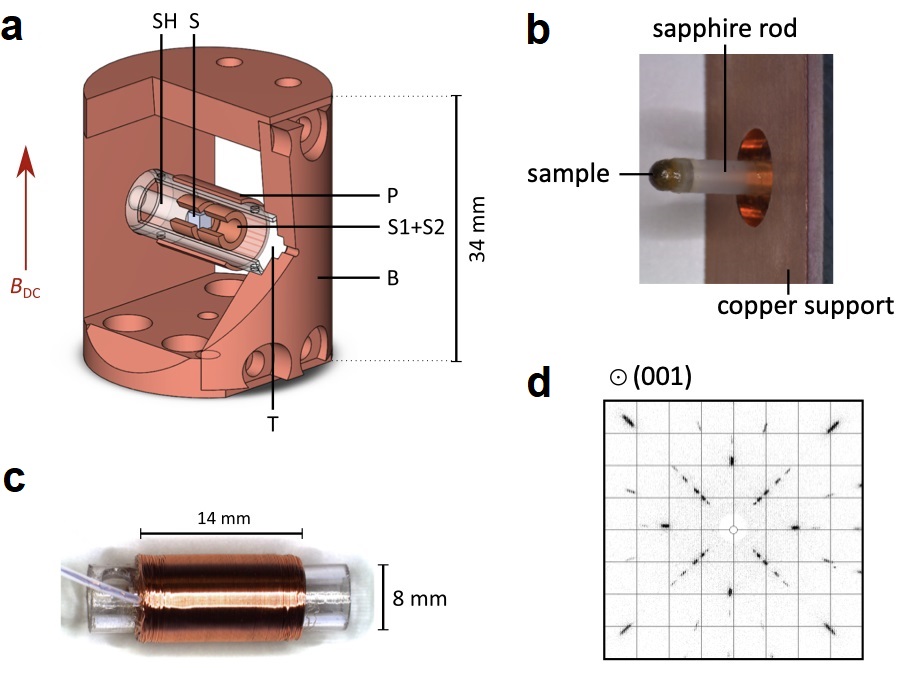}}
\linespread{1.0}\selectfont{}
\caption{\raggedright
{\bf $|$ Miniature AC susceptometer and single-crystal {\lhf} used in our study.}
\textbf{a}, Compact susceptometer for studies under transverse-field geometries at very low temperatures; picture reproduced from Ref.\,\onlinecite{2019_Rucker_RSI}.  A primary coil (P) and a balanced pair of secondaries (S1, S2) are placed on the outside and inside of a the sapphire tube, respectively. The sample (S) is mounted on a sapphire rod placed inside the secondaries. The assembly is rigidly mounted inside a body (B) made of high-purity Cu which provides excellent thermal anchoring. 
\textbf{b}, Spherical {\lhf} single-crystal as attached to the sapphire rod. 
\textbf{c}, Photograph of the sapphire tube and primary; picture reproduced from Ref.\,\onlinecite{2019_Rucker_RSI}.
\textbf{d}, Laue x-ray diffraction pattern of the {\lhf} single-crystal along the cylinder axis of the sapphire rod, confirming excellent alignment along the $c$-axis, i.e., the easy magnetic axis.
%\newline
%label: fig:EDI1
}
\label{fig:EDI1}
\end{figure*}

%%%%%%%%%%%%%%%%%%%%%%%%%%%%%%%%%%%
\clearpage \thispagestyle{empty}

\begin{figure*}[h]
%	\centering
	\centerline{\includegraphics[width=1.0\textwidth,clip=]{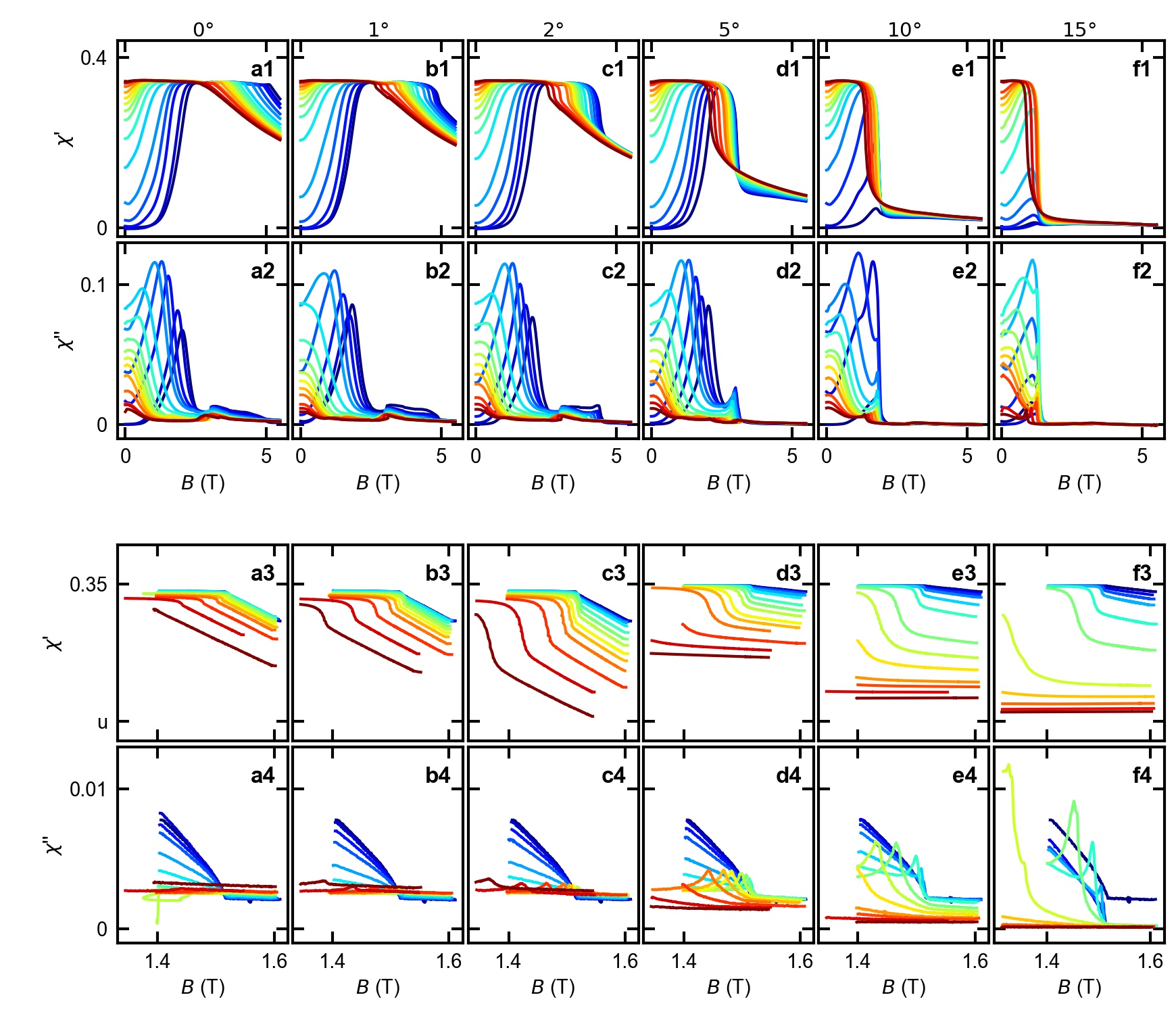}}
\linespread{1.0}\selectfont{}
\caption{\raggedright
{\bf $|$ Real and imaginary part of the transverse susceptibility recorded experimentally.} Data shown in each panel correspond to the value of $\phi$ stated at the top of each column of panels. The color coding of data recorded as a function of magnetic field at fixed temperature changes from blue to red going from low to high temperatures, respectively. The color coding of data recorded as a function of temperature at fixed magnetic field vary from blue to red  going from low field to high magnetic fields, respectively. Specific values of all parameters are summarized in Extended Data Table\,\ref{EDI:Theory:tab1}.
\textbf{a1} through \textbf{f1} (first row), Real part of the transverse susceptibility as a function of magnetic field  for different temperatures.
\textbf{a2} through \textbf{f2} (second row), Imaginary part of the transverse susceptibility as a function of magnetic field  for different temperatures.
\textbf{a1} through \textbf{f3} (third row), Real part of the transverse susceptibility as a function of temperature for different magnetic fields. Note that a different vertical scale was used for better visibility, where $u=0.28$ for panels \textbf{a3}, \textbf{b3}, and \textbf{c3}, and $u=0$ for panels \textbf{d3}, \textbf{e3}, and \textbf{f3}.
\textbf{a4} through \textbf{f4} (bottom row), Imaginary part of the transverse susceptibility as a function of temperature for different magnetic fields.
%\newline
%label: fig:EDI2
}
\label{fig:EDI2}
\end{figure*}

%%%%%%%%%%%%%%%%%%%%%%%%%%%%%%%%%%%
\clearpage \thispagestyle{empty}

\setcounter{figure}{2}
\captionsetup[figure]{labelfont={bf},name={Extended Data Fig.},labelsep=space}

\begin{figure*}[h]
%	\centering
	\centerline{\includegraphics[width=0.95\textwidth,clip=]{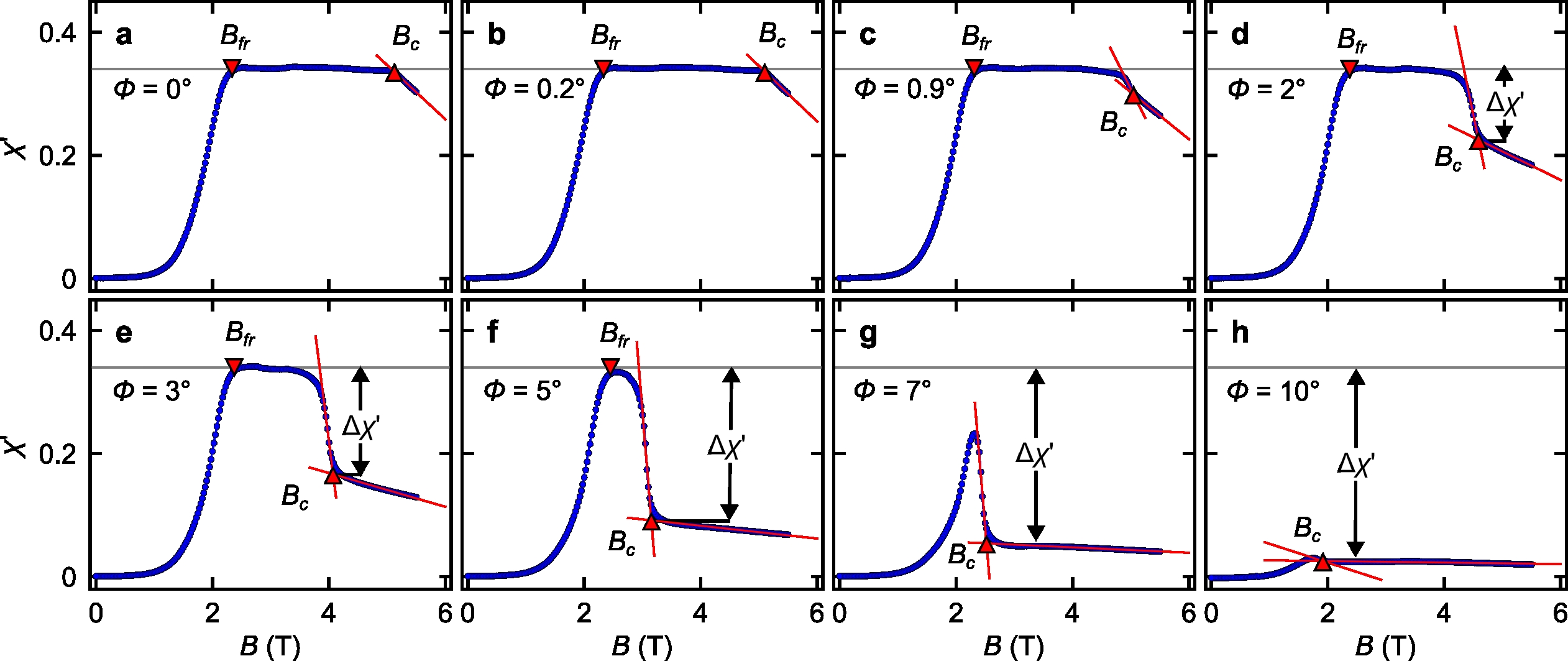}}
\linespread{1.0}\selectfont{}
\caption{\raggedright
{\bf $|$ Key features in the susceptibility.}
Definition of the critical field $B_c$, the freezing field, $B_{\rm fr}$, and the discontinuous change of the susceptibility at the transition, $\Delta\chi'$, for various tilt angles, $\phi$.
The critical field, $B_c$, represents the crossing point between a linear regression of the susceptibility in the paramagnetic state, and a linear regression of the fast rise of the susceptibility in the ferromagnetic state. Results reported in our manuscript are insensitive to the precise definition of $B_c$.
The freezing field, $B_{\rm fr}$, marks the onset of the decrease of the plateau of $\chi'$ under decreasing field, where the plateau is only fully developed for $\phi<5^{\circ}$.
The change of the susceptibility, $\Delta\chi'$, across the transition at $B_c$ represents the difference of the susceptibility at $B_c$ and the susceptibility of the plateau deep in the ordered state. As $B_c$ approaches $B_{\rm fr}$ for increasing $\phi$, the susceptibility of the plateau at low values of $\phi$ is used to determine $\Delta\chi'$.
%
%\newline
%label: fig:EDI8
}
\label{fig:EDI8}
\end{figure*}

%%%%%%%%%%%%%%%%%%%%%%%%%%%%%%%%%%%
\clearpage \thispagestyle{empty}

\begin{figure*}[h]
%	\centering
	\centerline{\includegraphics[width=0.8\textwidth,clip=]{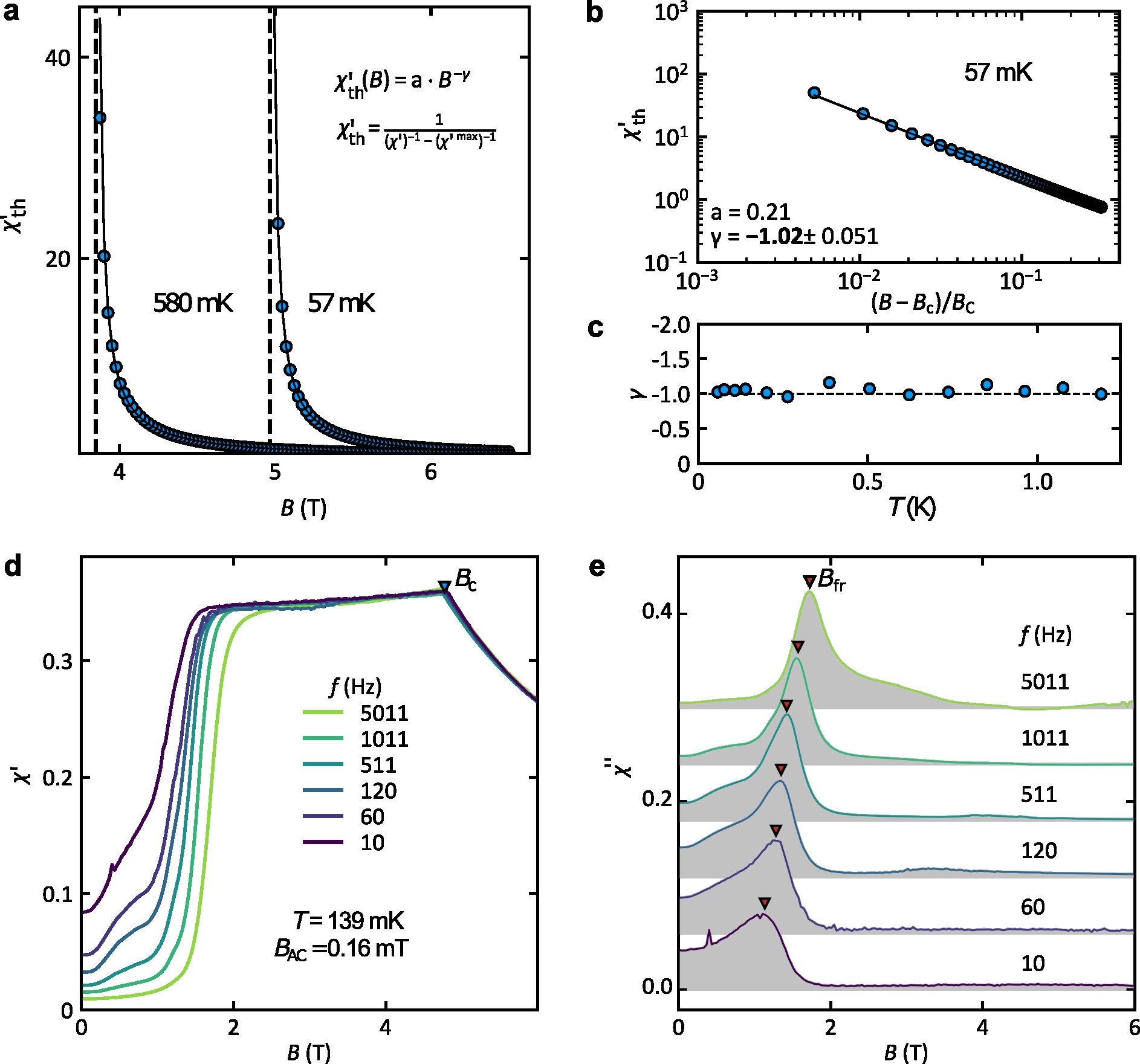}}
\linespread{1.0}\selectfont{}
\caption{\raggedright
{\bf $|$ Critical behaviour of the transverse susceptibility and frequency dependence of the freezing behaviour at low temperatures and low magnetic fields.} Data shown were recorded for $\phi=0$, i.e., ideal transverse field orientation.
\textbf{a}, Typical paramagnetic contribution of the real part of the transverse ac susceptibility, $\chi'_{\rm th}$, of {\lhf} for two selected temperatures. Contributions arising from the magnetic domains below $B_c$, denoted $\chi'^{\rm max}$, were subtracted from the measured susceptibility, $\chi'$.  The magnetic field dependence may be fitted by a power law dependence in excellent agreement with the literature\cite{1996_Bitko_PRL}.
\textbf{b}, Depiction of $\chi'_{\rm th}$ as a function of reduced magnetic field on a logarithmic scale at $57\,{\rm mK}$. Data points correspond to those shown in Ref.\,\onlinecite{2019_Rucker_RSI}. The critical exponent $\gamma=-1.02\pm0.0051$ is in excellent agreement with mean-field behaviour and the literature\cite{1996_Bitko_PRL}.
\textbf{c}, Critical exponent $\gamma$ at $B_c$ as a function of temperature. Along the phase boundary up to $\sim 1.2\,{\rm K}$ the same mean-field exponent, $\gamma\approx1$, is observed.
\textbf{d}, Real part of the transverse susceptibility, $\chi'$, as a function of magnetic field at $T = 139\,{\rm mK}$ for excitation frequencies between 10\,Hz and 5011\,Hz. With increasing frequency the onset of the freezing shifts to higher magnetic fields characteristic of a slow process.
\textbf{e}, Imaginary-part of the transverse susceptibility, $\chi''$, as a function of magnetic field at $T = 139\,{\rm mK}$ for various frequencies. $\chi''$ shows a strong peak as marked by the red triangles characteristic of dissipation at low magnetic fields and for all frequencies measured. With increasing frequency the peak shifts to higher magnetic fields characteristic of a slow process. Curves have been shifted by a constant for clarity.
%\newline
%label: fig:EDI3
}
\label{fig:EDI3}
\end{figure*}

%%%%%%%%%%%%%%%%%%%%%%%%%%%%%%%%%%%
\clearpage \thispagestyle{empty}

\begin{figure*}[h]
%	\centering
	\centerline{\includegraphics[width=0.8\textwidth,clip=]{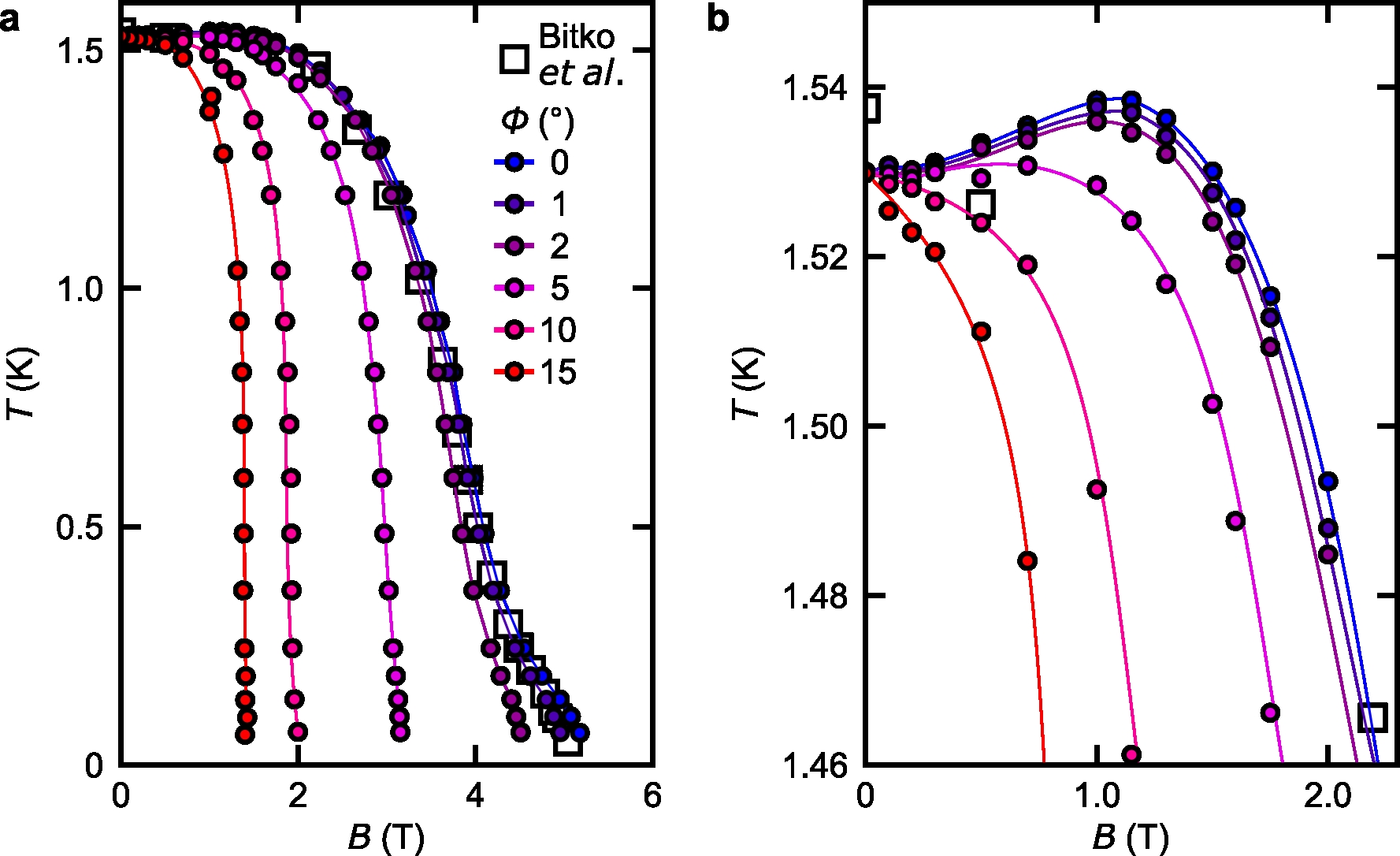}}
\linespread{1.0}\selectfont{}
\caption{\raggedright
{\bf $|$ Temperature versus magnetic field phase diagram for various field orientations $\phi$.}
A spherically shaped sample was measured to ensure uniform demagnetizing fields. \textbf{a}, Magnetic phase diagram of single-crystal {\lhf} for different field angles $\phi$. Data are identical to those shown in Fig.\,\ref{fig:2}\,\textbf{e} through \ref{fig:2}\,\textbf{i}. Data for $\phi=0$ are in excellent agreement with those reported by Bitko et al. in Ref.\,\onlinecite{1996_Bitko_PRL} are depicted by squares.
\textbf{b}, Close-up view of the magnetic phase diagram in the vicinity of the Curie temperature for small fields. The data suggest a tiny regime of reentrant behaviour as a function of magnetic field for $\phi=0$. Great care was exercised to correct  the small magnetoresistance of the RuO temperature sensor at low fields.\cite{2018_Rucker_PhD} Small differences of $T_c$ between our data and those reported in Ref.\,\onlinecite{1996_Bitko_PRL} may be attributed to small quantitative differences of thermometer calibration.
%\newline
%label: fig:EDI7
}
\label{fig:EDI7}
\end{figure*}

%%%%%%%%%%%%%%%%%%%%%%%%%%%%%%%%%%%
\clearpage \thispagestyle{empty}

\begin{figure*}[h]
%	\centering
	\centerline{\includegraphics[width=0.8\textwidth,clip=]{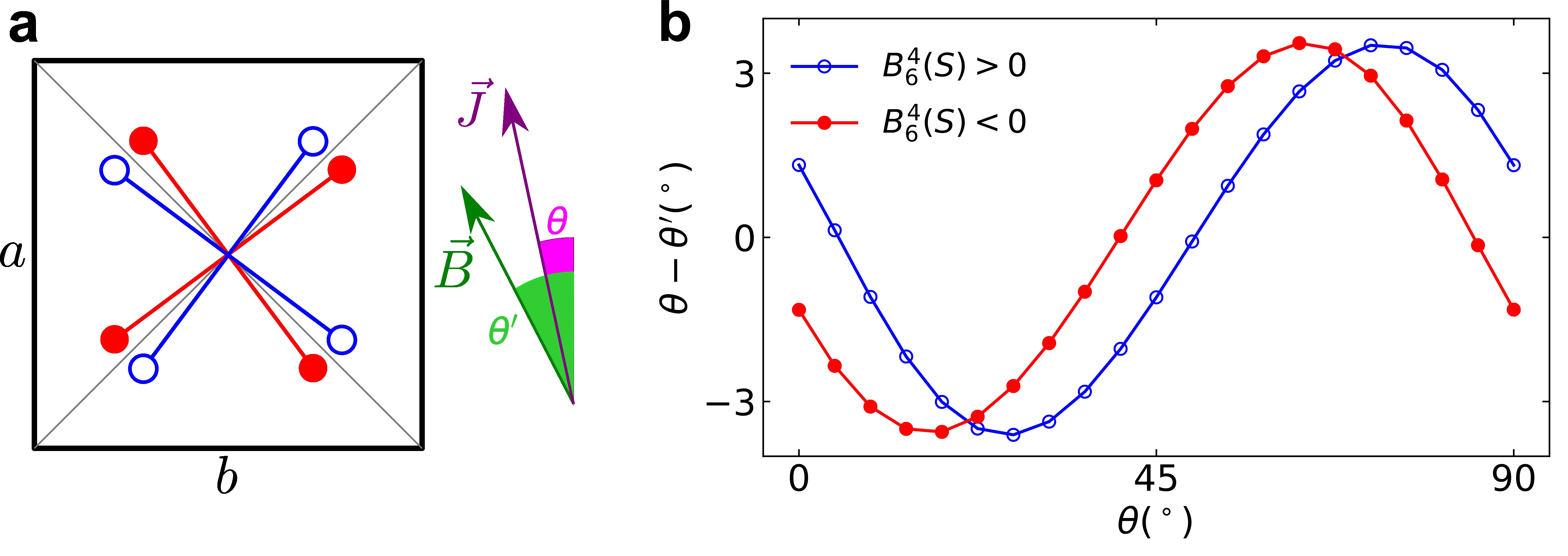}}
\linespread{1.0}\selectfont{}
\caption{\raggedright
{\bf $|$ Misalignment between the transverse field $\vec B$ ($\phi=0$) and the electronic moment $\vec J$ in the plane perpendicular to the easy axis at $B=6$\,T as calculated from the microscopic model Eq.\,\ref{eq:hmic1}.}
The symmetry of LiHoF$_4$ allows two configurations of the F$^-$ ions in the hard plane, which lead to different signs of the crystal field coefficient $B^4_6(S)$.\cite{2007_Ronnow_PRB} Due to this small symmetry-breaking crystal field the projection of the electronic moment onto the hard plane is not perfectly oriented along the applied magnetic field, but instead shows a small misalignment $\theta - \theta^\prime$. Thus even if the field is applied along a high-symmetry direction such as $\theta=0$, $\vec B = (B,0,0)^T$ the  electronic moment has a finite component perpendicular to the field axis $\theta^\prime \neq 0$, $J^y \neq 0$. Therefore the only spin symmetry that is not already broken in the disordered state at $B>B_c$ is the Ising symmetry $J^z \rightarrow -J^z$.
\textbf{a}, Qualitative sketch of the two equivalent configurations of the F$^-$ ions (circles) projected onto the hard magnetic plane as depicted in red and blue shading, resulting in opposite sign of the crystal field parameter $B_6^4(S)$. 
\textbf{b}, Misalignment angle $\theta-\theta'$ between the electronic moment $\vec{J}$ and the magnetic field direction, $\theta$, within the hard magnetic plane as a function the orientation of the field component in the hard plane.
%
%
%\newline
%label: fig:EDI6
}
\label{fig:EDI10}
\end{figure*}

%%%%%%%%%%%%%%%%%%%%%%%%%%%%%%%%%%%
\clearpage \thispagestyle{empty}

\begin{figure*}[h]
%	\centering
	\centerline{\includegraphics[width=0.95\textwidth,clip=]{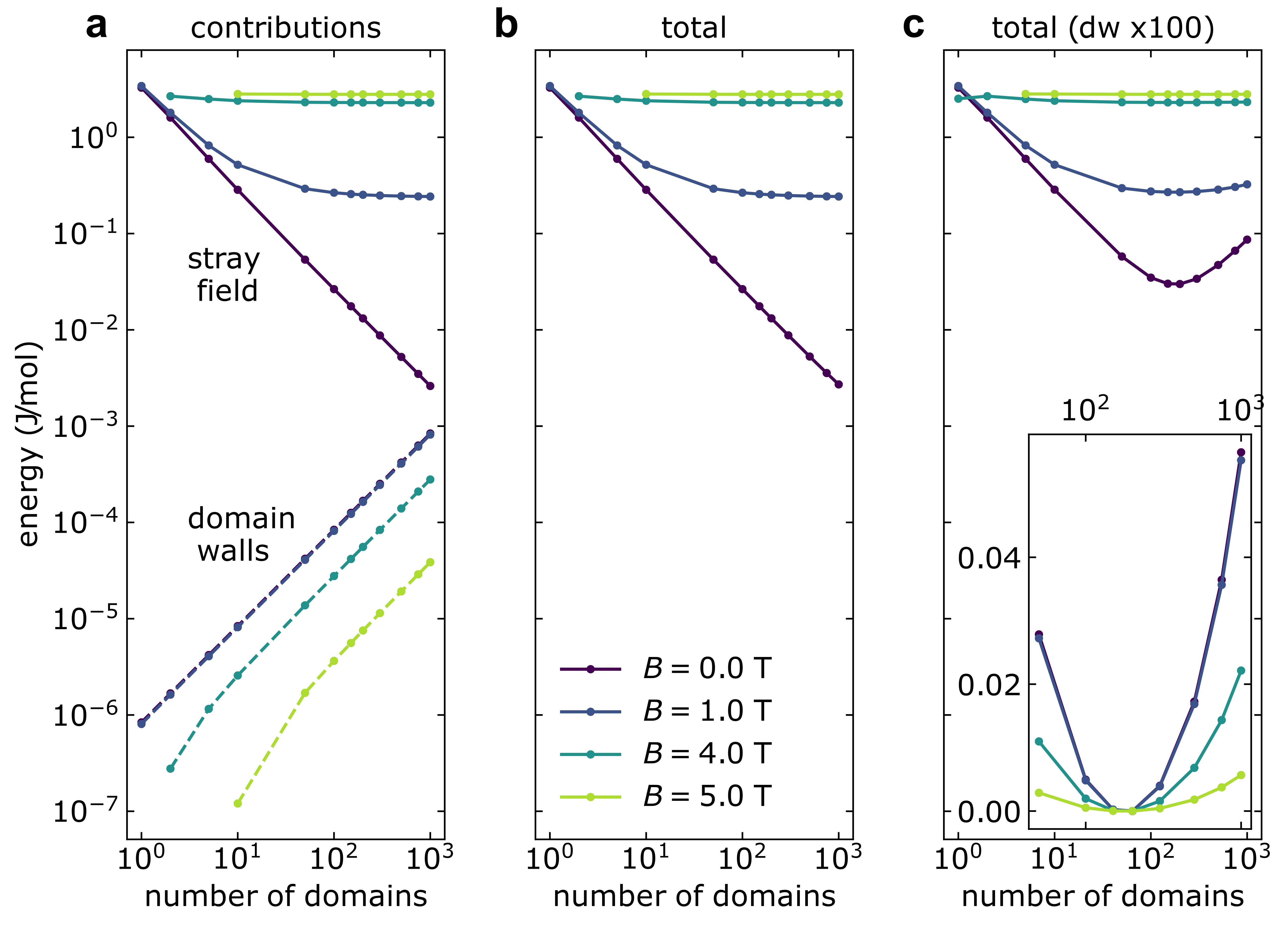}}
\linespread{1.0}\selectfont{}
\caption{\raggedright
{\bf $|$ Energy due to domains $E_{\rm dom}$ as a function of the number of domains, $N$, at various magnetic fields.}
The behaviour observed here implies that the theoretical analysis does not depend on the precise choice of the domain structure. Moreover, it allows to motivate the choice of parameters selected for the numerical evaluation. For details see Supplementary Note \,\ref{sec:energy}.
\textbf{a}, Individual contributions to $E_{\rm dom}$ as a function of $N$ at zero field, namely stray field energy and energy of the domain walls. The ideal number of domains is determined by the competition between these two terms, scaling roughly as $1/N$ and $N$ at B=0.
\textbf{b}, Total $E_{\rm dom}/M$ as a function of $N$. With increasing field strength the stray field contribution increases drastically, whereas the contributions by the domain walls decreases.
The full domain energy near the QCP is therefore dominated by stray-field contributions and only weakly dependent on $N$. The minimal energy and thereby the optimal $N$ is reached at $N>1000$.
\textbf{c}, Total $E_{\rm dom}/M$ as a function of $N$ when hypothetically assuming a domain wall energy that is $100$ times larger than realistic. This shifts the optimal $N$ to $80$. The minimum becomes shallow at large fields and therefore deviations from the optimal $N$ only lead to minor quantitative changes.
%
%\newline
%label: fig:EDI6
}
\label{fig:EDI9}
\end{figure*}

%%%%%%%%%%%%%%%%%%%%%%%%%%%%%%%%%%%
\clearpage \thispagestyle{empty}

\begin{figure*}[h]
%	\centering
	\centerline{\includegraphics[width=0.95\textwidth,clip=]{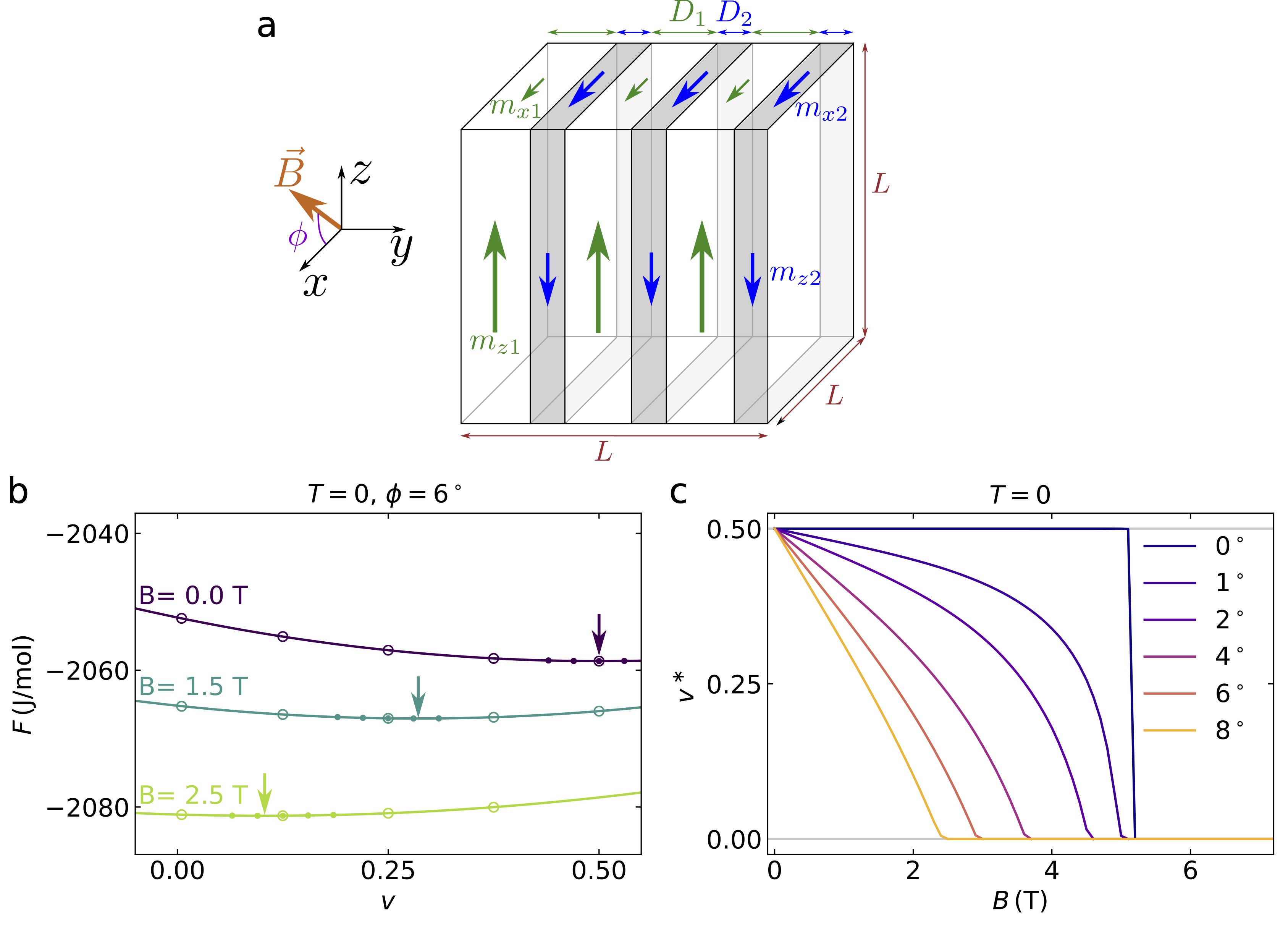}}
\linespread{1.0}\selectfont{}
\caption{\raggedright
{\bf $|$ Evolution of magnetic domains under magnetic field and various field directions $\phi$}.
\textbf{a}, Domain structure assumed in our model, where planar sheets are stacked along the $y$-axis. The spontaneous and the field-induced magnetization are oriented along the $z$- and the $x$-axis, respectively. The magnetic field is applied in the $zx$-plane. $D_{1}$ and $D_{2}$ denote the widths of the up and the down domains, respectively. The up and the down states represent the majority and minority domains, respectively.
\textbf{b}, Free energy $F$ in units K as a function of domain ratio $v=D_2/(D_1+D_2)$. The dashed vertical lines mark values of $v$, where $v^*$ denotes the stable configuration.
\textbf{c}, Optimal domain ratio $v^*$ of the minimum of $F$ as a function of magnetic field for different tilt angles $\phi$.
%\newline
%label: fig:EDI6
}
\label{fig:EDI6}
\end{figure*}

%%%%%%%%%%%%%%%%%%%%%%%%%%%%%%%%%%%
\clearpage \thispagestyle{empty}

\begin{figure*}[h]
%	\centering
	\centerline{\includegraphics[width=0.75\textwidth,clip=]{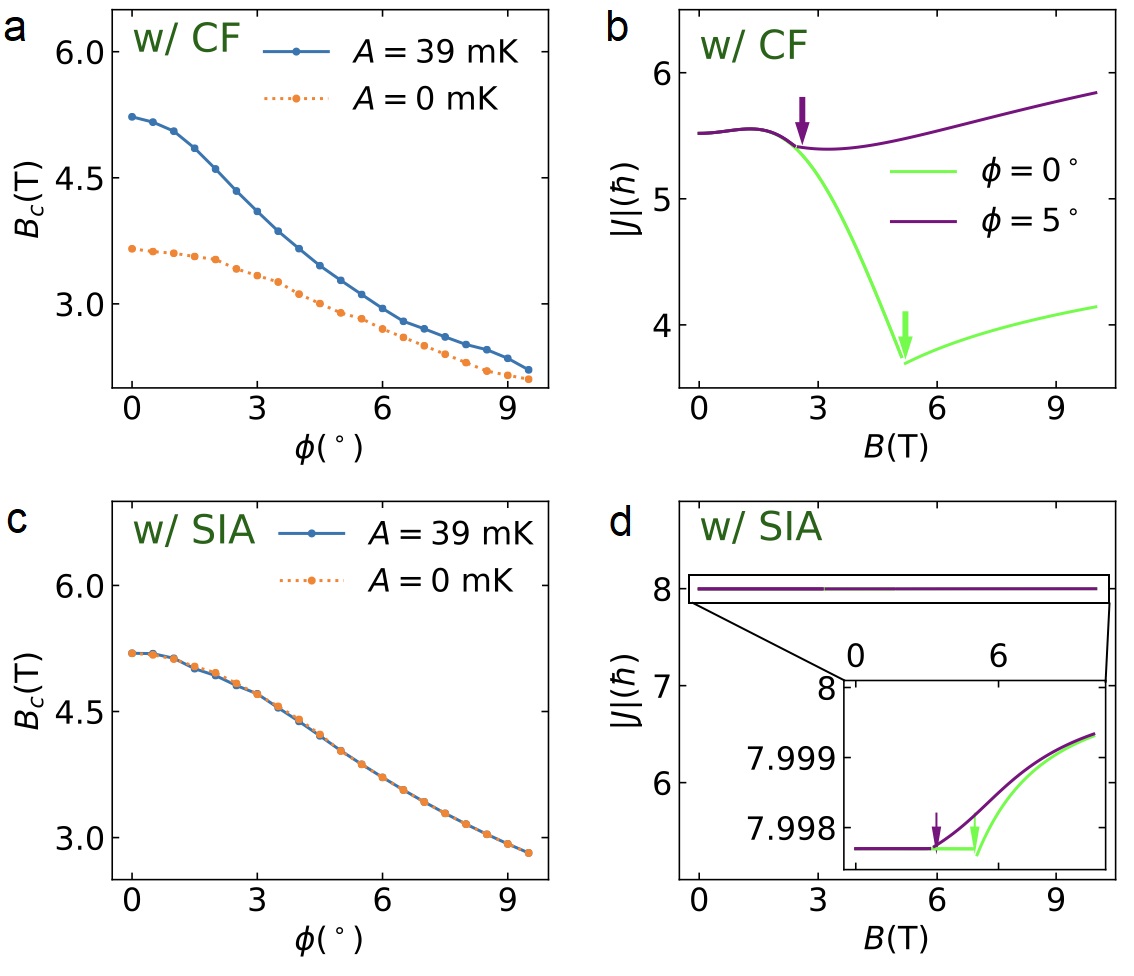}}
\linespread{1.0}\selectfont{}
\caption{\raggedright
{\bf $|$ Interplay of hyperfine coupling with the non-Kramers ground state at zero temperature.} Treatment of the Ising anisotropy in terms of the non-Kramers full crystal field terms is denoted (CF). Treatment of the Ising anisotropy in terms of a simple single-ion anisotropy acting on a Kramers moment is denoted (SIA).
\textbf{a}, Calculated critical field, $B_c$, as a function of field orientation, $\phi$, with and without hyperfine coupling $A$. The full set of crystal-field terms, $\vcf$, is taken into account. $B_c$ strongly increases due to the hyperfine coupling because the electronic moment of the Ho ion, $|J| = \sqrt{\langle {J^x}^2 + {J^y}^2 + {J^z}^2 \rangle}$, increases substantially in the ordered phase as shown in panel \textbf{b}. In turn, the Ho moment profits more strongly from the hyperfine coupling to anti-aligned nuclear spins. As $\phi$ increases, this effect is rapidly suppressed due to the crystal field terms. This accounts for the rapid decrease of $B_c$ as a function of increasing $\phi$ and the suppression of the additional increases of $B_c$ observed for $\phi=0$.
\textbf{c}, Calculated critical field, $B_c$, as a function of field orientation, $\phi$, with and without hyperfine coupling $A$, where the Ising character is accounted for by a SIA acting on a Kramers moment. The critical field does not exhibit a substantial dependence on the hyperfine coupling and the sensitivity of $B_c$ to changes of $\phi$ is much reduced, because $|J|$ is essentially field-independent as shown in panel \textbf{d}. The tiny variation of $|J|$ under these conditions, highlighted in the inset, reflects the hyperfine-induced entanglement of the electronic moment with the nuclear spin.
%
%\newline
%label: fig:EDI5
}
\label{fig:EDI5}
\end{figure*}

%%%%%%%%%%%%%%%%%%%%%%%%%%%%%%%%%%%

%%%%%%%%%%%%%%%%%%%%%%%%%%%%%%%%%%%

\clearpage \thispagestyle{empty}

\setcounter{figure}{0}
\captionsetup[figure]{labelfont={bf},name={Extended Data Table},labelsep=space}

%\setcounter{table}{0}
%\captionsetup[table]{labelfont={bf},name={Extended Data Table},labelsep=space}
%\input{_DisplayItems/EDI-tab1-v1}

\begin{figure*}[h]
%	\centering
\linespread{1.0}\selectfont{}
\caption{\raggedright
\textbf{$|$ Parameters of data shown in Fig.\,\ref{fig:2} and Extended Data Fig.\,\ref{fig:EDI2}.} Parameter ranges of temperature and field sweeps and associated values of the magnetic field, temperature, and tilt angle during the sweeps. See methods for information on accuracies of values stated. 
}
\centerline{\includegraphics[width=1.0\textwidth,clip=]{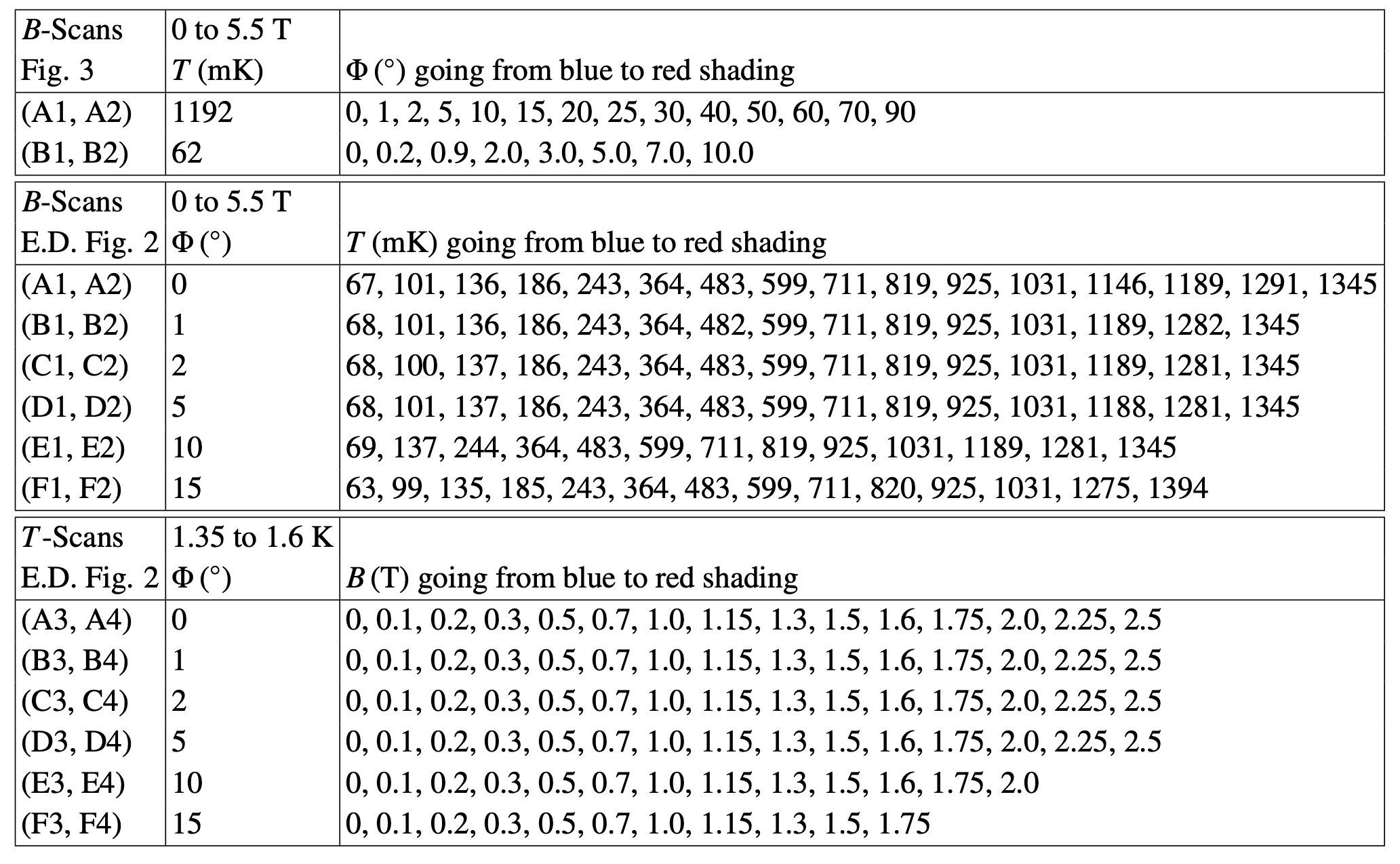}}
\label{EDI:Theory:tab1}
\end{figure*}

%%%%%%%%%%%%%%%%%%%%%%%%%%%%%%%%%%%
%\end{document}

%%%%%%%%%%%%%%%%%%%%%%%%%%%%%%%%%%%
%\clearpage \thispagestyle{empty}

%%%%%%%%%%%%%%%%%%%%%%%%%%%%%%%%%%
\clearpage \thispagestyle{empty}
%%%%%%%%%%%%%%%%%%%%%%%%%%%%%%%%%%

\setcounter{page}{1}

\newpage
\newpage
\section*{Supplementary Information}

\setcounter{section}{0}
\renewcommand{\thesection}{S\arabic{section}}

In these Supplementary Notes we present additional information on the theoretical derivations and numerical evaluation of our model. The details reported are intended to be pedagogical. The Supplementary Notes are organized in eight sections.
We begin in Supplementary Note~\ref{sec:theory} with comments on various aspects of the theoretical analysis, notably possible scenarios of the line of phase transitions under tilted fields, and a review of theoretical models of LiHoF$_4$ reported in the literature so far.
In Supplementary Note~\ref{sec:landau} we introduce the phenomenological theoretical framework of the transverse field quantum phase transition under tilted magnetic fields.
This is followed in Supplementary Note~\ref{sec:microscopic} by an account of the microscopic Hamiltonian for LiHoF$_4$, notably the Ho$^{+3}$ ions, taking into account the crystalline electric field (CEF) potential, Zeeman terms of the tilted field, hyperfine coupling to nuclear spins and interaction between electronic spins.
In Supplementary Note~\ref{sec:meanfield} we present the mean-field approximation of the microscopic model, involving the diagonalization of the $17\times8$ matrix.
We then move on to review the literature on magnetic domains and domain patterns in Supplementary Note~\ref{sec:domains} in order to motivate our choice of sheet-like domains considered in our model.
In Supplementary Note~\ref{sec:interactions} we deduce an exact expression for the stray-field energy for a given domain pattern and find that it can be rewritten in terms of effective interactions between average magnetization components in the different domains in the spirit of an antiferromagnetic coupling.
We then proceed to combine the microscopic interactions and the domain energy into a joint mean-field description described in Supplementary Note~\ref{sec:combined}.
In a first assessment of the combined model, presented in Supplementary Note~\ref{sec:energy}, we discuss the energy of the stray field and the domain walls as a function of domain number. This shows that our results are insensitive to the precise choice of domain shapes and validates the assumptions we make.
Supplementary Note~\ref{sec:numerical}, finally, describes how the full Hamiltonian is solved in a mean-field approximation where the domain pattern is optimized in the spirit of a variational ansatz.
Here we address in particular changes of the phase diagrams in the presence of domains, as well as the intricate interplay of the non-Kramers character of the Ho moments with the hyperfine coupling.

%%%%%%%%%%%%%%%%%%%%%%%%%%%%%%%%%%
\section{Aspects of the theoretical analysis}
\label{sec:theory}

We report the experimental observation of a well-defined line of phase transitions that emanates under tilted magnetic fields from the text-book example of a transverse-field quantum critical point. The observation of the phase transition under tilted fields raises the question for the existence and nature of the underlying symmetry breaking. Close inspection of the well understood microscopic spin Hamiltonian of LiHoF$_4$ reveals that its spin symmetry is fully broken in the presence of an applied magnetic field tilted away from the hard plane. A subtlety concerns here that the atomic positions of the F$^-$ ions imply also a symmetry breaking within the hard plane. Thus, even under perfectly transverse fields the paramagnetic phase displays a tiny magnetization component perpendicular to the applied field as illustrated in Extended Data Fig.\,\ref{fig:EDI10}. Representing a small effect, this does not question the validity of the account of the TF-QCP as such.

Given that the spin symmetries of the microscopic Hamiltonian under magnetic field are fully broken, only a few putative mechanisms exist of the symmetry breaking. Appreciating that the formation of domains has long been known in LiHoF$_4$, the simplest mechanism is a breaking of translation symmetry associated with the formation of magnetic domains.
Alternative causes of the symmetry breaking comprise some form of microscopic antiferromagnetic or polar order, a transition of the electronic structure (say a metal-insulator transition), or a transition of the crystal structure. These options are exceedingly unlikely. Namely, there is no evidence suggesting additional magnetic interactions that may cause a transition of the microscopic magnetic structure even under tiny tilted magnetic fields. Further, the crystal structure has full inversion symmetry and polar degrees of freedom are not expected to exist. Also, the electronic structure is very well understood with a large band gap far from any instabilities. Last but not least, the crystal structure is known to be very stable under substitutional doping as well as for the entire series of iso-structural Li-R-F$_4$ compounds, where R is a rare-earth. 

Hence, the observation of a line of phase transitions that emerges continuously from the TF-QCP and evolves monotonically as a function of field direction appears to be very difficult to reconcile with the four mechanisms just mentioned. Moreover, the field and temperature dependence of the susceptibility evolves monotonically between the well-understood behavior at the TF-QCP under transverse fields to the characteristics of a conventional hysteresis loop for field along the easy axis. In turn, the formation of magnetic domains appears to represent the only plausible explanation of the line of phase transitions under tilted fields.

At this point it is instructive to summarize concepts and models from the literature used to describe transverse-field Ising systems in general and LiHoF$_4$ in particular. These set the stage for the account of ferromagnetic domains under tilted fields. In general, the models can be distinguished according to their local degrees of freedom, the interactions between them and the treatment thereof, and whether or not mesoscopic inhomegeneities are taken into account.
A first group of theoretical studies explored general aspects of Ising ferromagnetism. This includes the properties of branched domain structures for Ising ferromagnets.\cite{Gabay84, Gabay85, 1986_Gabay_PRB}  Using an effective spin $\nicefrac{1}{2}$ model without hyperfine interactions the thermodynamic signatures and domain structure at zero magnetic field was found to be in excellent agreement with experiment.\cite{2009_Biltmo_EPL} Focussing on the possible existence of domain wall roughening by means of the effective spin $\nicefrac{1}{2}$ a renormalization group analysis no domain wall roughening for long-range interactions at zero and high temperature was observed.\cite{Mias05} Even an accurate estimate of the demagnetization factors was reported recently, however, without discussion of the magnetic field dependence and quantum criticality.\cite{2020_Twengstrom_PRB} 

Turning to specific models for LiHoF$_4$, the seminal work of Bitko et al. \cite{1996_Bitko_PRL} employed the complete Hamiltonian of the Ho$^{3+}$ with a $J=8$ electronic spin and $I=7/2$ nuclear spin in the full CEF scheme resulting in a $(17\times 8)$ matrix. Bitko et al evaluated this model on a mean-field level, using an  empirical demagnetization correction to correct for domains. A similar approach was pursued in studies of the collective excitations by Ronnow et al \cite{2005_Ronnow_Science, 2007_Ronnow_PRB, 2016_Kovacevic_PhysRevBa}, where the effects of fluctuations were included using a $1/z$ expansion in an effective medium approach as first introduced for HoF$_3$.\cite{1994_Jensen_PRB}

To reduce computational complexity, the work by Chakraborty et al. \cite{2004_Chakraborty_PRB} developed an effective spin $\nicefrac{1}{2}$ model of the non-Kramers ground state, keeping the $I=7/2$ nuclear spins. A Clausius-Mosotti equation was used to determine the effects of magnetic domains, incorporating the effects in terms of a renormalization of the couplings constants. Evaluating their model by means of quantum Monte Carlo simulations, the hyperfine coupling was taken into account a posteriori in terms of a renormalization of the applied field. Taking this model one step further, McKenzie and Stamp \cite{2018_McKenzie_PRB} recently solved the effective spin $\nicefrac{1}{2}$ model using mean-field theory and RPA. To interpret microwave spectroscopy of electronuclear modes near quantum criticality another variant of the effective spin $\nicefrac{1}{2}$ model was used, where the effects of ferromagnetic domains was included by means of an effective demagnetization factor while not considering the domain structure as such.\cite{2021_Libersky_arxiv} Finally, several studies explored also the validity of the effective spin $\nicefrac{1}{2}$ model in doped systems, where random fields are dominant.\cite{2005_Schechter_PhysRevLetta, 2008_Schechter_PhysRevBa, 2008_Tabei_PhysRevBb}

For the description of LiHoF$_4$ in tilted fields, it turns out to be crucial to include both microscopic and mesoscopic degrees of freedom \cite{2021_Eisenlohr_PhD}, with the latter referring to the sizes of magnetic domains. Aiming at a quantitatively accurate description, we include the complete a $(17\times 8)$-dimensional local Hilbert space, as in earlier works,\cite{1996_Bitko_PRL} and combine this with the magnetostatics of an arrangement of magnetic domains which is assumed to be periodic for simplicity. We treat the dominantly dipolar interaction in a mean-field approximation, but it is important to keep distinct mean fields for up and down domains to account for the explicit symmetry breaking by the field tilt. Hence, combining such a variable domain description with an accurate microscopic mean-field theory is the new aspect of our approach.

%%%%%%%%%%%%%%%%%%%%%%%%%%%%%%%%%%
\section{Landau theory}
\label{sec:landau}

The simplest phenomenological description of a transverse-field Ising quantum phase transition in Landau theory is by means of a scalar order parameter $M$ with Ising symmetry. The (bare) order-parameter mass is given by $(h-h_c)$ for perfectly transverse field $h$, such that an ordered phase is realized for $h<h_c$. Accounting for tilting of the field by an angle $\phi$ yields a Landau functional of the form
\begin{equation}
F(M)= \frac{a}{4} M^4 + \frac{1}{2}\left(  h \cos\phi - h_c \right) M^2 + M h \sin\phi .
\end{equation}
where $h\sin\phi$ represents a field conjugate to the order parameter. For $\phi=0$ the transition at $h_c$ displays mean-field exponents, i.e., $M \propto (h-h_c)^\beta$ and $M^\delta \propto h_c \phi$.

%\todo{add sentence on parameters and field}

The presence of a longitudinal field at finite $\phi$ renders $M$ non-zero for any order-parameter mass and hence destroys the field-driven phase transition. The transition turns into a crossover, whose location $h^\ast$ may be defined via the maximum of the susceptibility, $\chi_h = \partial M/\partial h$, as a function of $h$. A straightforward calculation for small $\phi$ yields
\begin{equation}
h_c-h^\ast \propto \phi^{2/3}
\end{equation}
The general result is expected to be $\phi^{1/(\beta \delta)}$ deduced from the critical power laws.
Our explicit microscopic calculation for {\lhf}, described in more detail below, indeed yields a crossover at $B_c-B^\ast \propto \phi^{2/3}$ if domain effects are neglected, see Fig.\,\ref{fig:3}\,\textbf{d}.

In stark contrast, the experiment measures qualitatively different behavior. Instead of a crossover, a sharp transition is observed, with the critical field approximately shifting proportional to $\phi^2$. The existence of a sharp thermodynamic transition at finite tilt angles is clearly beyond a single-component order-parameter description in the spirit of Landau theory.

%%%%%%%%%%%%%%%%%%%%%%%%%%%%%%%%%%
\section{Microscopic modelling}
\label{sec:microscopic}

A single Ho$^{3+}$ ion in a magnetic field applied under an angle $\phi$ with respect to the hard axis, $\vec B= (B \cos \phi, 0, B \sin \phi)$,  and subject to a crystal field of $S_4$ symmetry, which in LiHoF$_4$ is created by the neighboring $F^-$ ions \cite{2008_Tabei_PRB}, can be described by the Hamiltonian
\begin{equation}
H_{\rm ion} = \vcf(\vec J)  + A \vec J \cdot \vec I - \mu_B \vec B \cdot (g \vec J + g_N \vec I)
\end{equation}
with $J=8$ electronic moments and $I=7/2$ nuclear moments. The hyperfine coupling to the nuclear spins is assumed to be of Heisenberg type \cite{2018_McKenzie_PRB}. The associated material-specific strength of the hyperfine coupling for {\lhf} is $A=39$\,mK. The CEF term is given by
\begin{eqnarray}
\vcf(\vec J)
 = B_2^0 O_2^0 &+& B_4^0 O_4^0 + B_6^0 O_6^0 + B_4^4(C) O_4^4(C) \nonumber\\
 &+& B_4^4(S) O_4^4(S) + B_6^4(C) O_6^4(C) + B_6^4(S) O_6^4(S)
\end{eqnarray}
with the Stevens operators $O_l^m$ and the coefficients $B_l^m$ taken from Ref.~\onlinecite{2004_Chakraborty_PRB}. The Stevens operators are polynomials in $\vec J$ that encode the angular dependence, while the radial component is included by fitting their coefficients $B_l^m$ to experimental data. We note that $B_6^4(S)$ may be either positive or negative depending on the crystallographic position of the F$^-$ ions\cite{2007_Ronnow_PRB}, see Extended Data Fig.~\ref{fig:EDI10}.

The electronic Land\'{e} factor $g$ of a single Ho$^{3+}$ ion can be derived from the Wigner-Eckardt theorem to be $5/4$ \cite{2004_Chakraborty_PRB}. Small deviations from this single-ion value are expected in the extended crystal \cite{1975_Cooke_JPC,1975_Battison_JPC}. To achieve good agreement with experimental data we used $g=1.1$, i.e., $12\%$ lower than the single-ion value. Without this adjustment our results remain qualitatively unchanged, but there is a larger mismatch of $B_c/T_c$ compared to experiment. We note that the value of $g$ not only re-scales the magnetic field but also influences the stray-field energetics, see below.
For completeness we also include the nuclear Zeeman term with $g_N=1.5 \times 10^{-3}$, although its effect on observables is small because the behavior of the nuclear spins is dominated by the large hyperfine coupling.

Diagonalization of $\vcf$ in its $17$-dimensional electronic Hilbert space gives a low-lying non-Kramers doublet which is separated from the next CEF level by an energy gap of $11$\,K, Fig.\ref{fig:1}\,\textbf{c}. The ground-state doublet displays a large moment along the Ising axis, $\langle J^z\rangle \approx \pm 5.5$, but zero moment along $x$ and $y$. As a result, the application of a magnetic field along $x$ leads to the corresponding moment scaling as $\langle J^x \rangle \propto {B^x}^2$, Fig.\ref{fig:1}\,\textbf{d}, resulting from mixing with higher CEF levels.
In contrast to parts of the literature \cite{1996_Bitko_PRL, 2005_Ronnow_Science, 2005_Schechter_PhysRevLetta, 2004_Chakraborty_PRB, 2008_Tabei_PRB, 2018_McKenzie_PRB}  where the ground-state doublet was treated as a pseudospin $1/2$, we keep the full $17$-dimensional Hilbert space. This way we account fully for the non-Kramers physics in our calculation. This proves to be essential when describing the effects of an applied magnetic field in both the $z$ \textit{and} the $x$ direction.

Ferromagnetism in LiHoF$_4$ is driven by interactions between the electronic spins of different Ho$^{3+}$ ions, which involve both long-range dipolar and nearest-neighbor exchange interactions. For simplicity, we model this combination by a nearest-neighbor Heisenberg interaction $K$ \cite{2008_Tabei_PRB}, recalling that the dominant source of magnetic anisotropy is the CEF term.

Taken together, this leads to a microscopic Hamiltonian of electronic spins $\vec J$ and nuclear spins $\vec I$ of the form
\begin{equation}
\label{eq:hmic1}
H_{\rm mic}
=  - K \sum_{\langle ij\rangle} \vec J_{i} \cdot \vec J_{j} + \sum_i \big[V_{\rm CF}(\vec J_i) + A \vec J_i \cdot \vec I_i \big]  - \mu_B \vec B \cdot \sum_i (g \vec J_i + g_N \vec I_i),
\end{equation}
where $\langle ij\rangle$ runs over pairs of nearest neighbors. At this point the strength of the effective interaction reduces to a fit parameter. For good agreement with the experimental critical temperature $T_c$ we choose $K=14.5$\,mK, which is consistent with the order of magnitude of values estimated for the dipolar and exchange couplings in LiHoF$_4$. \cite{2018_McKenzie_PRB}

%%%%%%%%%%%%%%%%%%%%%%%%%%%%%%%%%%
\section{Mean-field approximation}
\label{sec:meanfield}

Due to the long-ranged nature of the dipolar interaction in {\lhf}, the effects of fluctuations are suppressed and a mean-field treatment appears justified. In particular, the upper critical dimension of the finite-temperature transition is $d_c^+=3$ -- as opposed to $4$ for short-ranged interactions -- such that the phase transition is of mean-field type both at $T=0$ and finite $T$ (the latter with logarithmic corrections).\cite{larkin69}

Within mean-field approximation, the Hamiltonian in Eq.~\eqref{eq:hmic1} reduces to a single-site problem
\begin{equation}
\label{eq:hmic2}
H_{\rm mic}^{\rm MF} =  - nK (\vec J \cdot \vec{\bar J} - \frac{\bar{\vec{J}}^2}{2}) + V_{\rm CF}(\vec J) + A \vec J \cdot \vec I - \mu_B \vec B \cdot (g \vec J + g_N \vec I),
\end{equation}
where $n=4$ represents the number of nearest neighbors. Solving $H_{\rm mic}^{\rm MF}$ amounts to the diagonalization of a $17\times 8$-dimensional matrix, supplemented by the self-consistency condition $\vec{\bar J} = \langle \vec J \rangle$.\cite{1996_Bitko_PRL} 

The resulting zero temperature phase diagram of this purely microscopic (i.e. single-domain) scenario is displayed in Fig.~\ref{fig:3}\,\textbf{d}. It features a quantum phase transition as function of applied field at $\phi=0$, i.e., for perfectly transverse field, and a crossover at a field $B^*$ under tilted fields $\phi\neq0$. The properties of this crossover are consistent with the results from Landau theory as outlined in Supplementary Information Note~\ref{sec:landau}. Namely, the susceptibility $\chi^{zB} = \partial \bar J^z / \partial B$ displays a maximum as function of applied field at $B*$, the location of which is marked by arrows in Fig.~\ref{fig:3}\,\textbf{d}. At small angles, we find $B_c-B^\ast \propto \phi^{2/3}$ as expected. However, the level of theoretical modelling is insufficient to explain the experimental observations. Instead the well-known fact that ferromagnets exhibit domains must be included.

%%%%%%%%%%%%%%%%%%%%%%%%%%%%%%%%%%
\section{Relevance of magnetic domains}
\label{sec:domains}

The formation and shape of magnetic domains in uniaxial (Ising) ferromagnets has been studied extensively starting with the seminal papers of Landau/Lifshitz and Kittel.\cite{1935_Landau_PZS, 1946_Kittel_PR, 1949_Kittel_RMP, 2000_Wolf_BJP, 2008_Hubert_Book} It has also long been appreciated that the shape of a given sample strongly influences the formation and shape of the domains.\cite{Luttinger46, Griffiths68, 2008_Hubert_Book} In platelets two remanent domain structures were observed subject to the thickness.\cite{1963_Gemperle_pss,1970_Kaczer_IEEE} First, the parallel-plate domain structure proposed by Landau and Lifshitz, where the easy-axis is normal to the platelets. Second, the honey-comb domain structure consisting of closely packed arrays of circular cylindrical domains. Detailed assessments revealed that the total free energy of the two domain structures differ by tenths of percent being essentially identical. Further studies revealed the presence of closure domains and branching at the surface of the samples that reduce the stray field energy.\cite{1988_Pommier_JdP, 1989_Meyer_EOMOMatAppl, Gabay84, Gabay85, 1986_Barker_JPC, 2014_Karci_RSI, 1986_Gabay_PRB}
Extensive studies explored the dynamical properties of domain structures and domain walls in Ising ferromagnets where a logarithmic rather than an exponential time dependence of relaxation processes was attributed to a wide distribution of barriers arising from the domain branching.\cite{1986_Gabay_PRB, 1990_Koetzler_PRL}

The domain structure of LiHoF$_4$ was first studied experimentally by means of Faraday rotation, finding evidence for stripe-like domains along the magnetic easy axis with a thickness of order $\sim 5\,\mu{\rm m}$. \cite{1975_Battison_JPC} Further studies analyzed the domain pattern in more detail for a slab-like geometry as a function of a longitudinal field at $T=1.3\,$K, where discontinuous transitions from a stripe pattern to a bubble pattern followed by the uniform state were reported. \cite{1988_Pommier_JdP, 1989_Meyer_EOMOMatAppl} Moreover, branching near the surface was observed, in agreement with theoretical results in dipolar Ising magnets. \cite{Gabay84, Gabay85} This is consistent with more recent studies using scanning Hall microscopy, which reported also the observation of substructures within the domains. In addition, these studies reported changes of the domain size as a function of temperature and transverse field they attributed to surface branching. \cite{2014_Karci_RSI, MAJorba}

% Chakrabothy: Clausius Mosotti; aim to explain mismatch of Bc und Tc;  no domain structure; no tilted fields; finite susceptibility;

Theoretical studies of LiHoF$_4$ addressing the long-ranged dipolar interaction employed an Ewald summation \cite{Ewald21} or the reaction-field method.\cite{Barker73} The former naturally leads to domain formation in Monte Carlo simulations with a large enough unit cell and appropriate boundary conditions.\cite{2009_Biltmo_EPL} In Ref.~\onlinecite{2009_Biltmo_EPL} a domain pattern of parallel sheets was found to be favorable at $T=0$ and $\vec B=0$, and the energy density of domain walls was estimated. The latter incorporates the presence of domains by considering an imaginary microscopic sphere in an effective field of surface charges.\cite{2004_Chakraborty_PRB} The structure of domain walls in transverse-field Ising models was investigated in Ref.~\onlinecite{Mias05} and long-ranged interactions were found to prevent domain-wall roughening.

In summary, due to the strong Ising anisotropy the domain patterns are comparatively simple in several important aspects. First, the two dominant domain patterns observed experimentally are energetically almost equivalent\cite{1970_Kaczer_IEEE,2000_Wolf_BJP, 1975_Battison_JPC, 1988_Pommier_JdP, 1989_Meyer_EOMOMatAppl, 2014_Karci_RSI, MAJorba}  suggesting that the precise choice is not important at the level of our results. This is consistent with the numerical evaluation of our model presented in Supplementary Note\,\ref{sec:energy} which justifies to focus on a scenario that is mathematically amenable. Second, the observation of branching is confined to the surface of samples as reported in the literature.\cite{1986_Barker_JPC, 1988_Pommier_JdP, 1989_Meyer_EOMOMatAppl, Gabay84, Gabay85, 1986_Gabay_PRB, 1990_Koetzler_PRL} It concerns a small volume fraction only that will not change the main conclusions of our study. Third, due to the strong Ising anisotropy roughening of the domain walls is not expected.\cite{1986_Gabay_PRB, 1990_Koetzler_PRL, Mias05} Taken together it appears therefore well-justified to account for the formation of domains by means of a simple antiferromagnetic arrangement on mesoscopic scales.

%%%%%%%%%%%%%%%%%%%%%%%%%%%%%%%%%%
\section{Effective interactions induced by domain formation}
\label{sec:interactions}

As argued above, and consistent with our mean-field approach, it proves to be sufficient to refrain from a detailed modelling of domain and sample shape.
We assume a cubic sample with base length $L$, with $L=0.005$\,m for the results shown, and a periodic arrangement of two types of sheetlike domains stacked along the $y$ direction, Extended Data Fig.~\ref{fig:EDI6}\,\textbf{a} (cf. Fig.\ref{fig:3}\,\textbf{f} and \ref{fig:3}\,\textbf{g}).
We note that an alternative domain arrangement with sheet-like domains along the $x$ instead of the $y$ direction is energetically unfavorable due to induced charges on the domain walls. We assume, further, a homogeneous magnetization within each domain.

The domain thicknesses are denoted $D_{1,2}$ and their magnetization densities $\vec m_{1,2}$, where the indizes 1 and 2 refer to the majority (up) and minority (down) population, respectively. At zero field we expect $\vec m_1 = -\vec m_2 \parallel z$ and $D_1=D_2$, reflecting the Ising symmetry. At finite field, the magnetizations will develop both $x$ and $z$ components, and the domain sizes can be different. Assuming $D_1\geq D_2$, we introduce the volume fraction of the minority domains as
\begin{equation}
v = \frac{D_2}{D_1+D_2} = \frac{D_2}{2D}
\end{equation}
where $D=L/(2N)$ is the average domain thickness and $N$ the number of domains of each type. In what follows below, the fraction $v$ will be treated as a variational parameter.

In general, the total stray field energy of a sample without volume charges ($ \nabla \vec m=0$) is given by
\begin{equation}
E_d  = \frac{\mu_0}{8\pi} \int d^2 r d^2 r^\prime \frac{\sigma(\vec r) \sigma(\vec r^\prime)}{|\vec r-\vec r^\prime|}
\end{equation}
where the surface charge $\sigma(\vec r)$ is given by $\sigma(\vec r) = \vec m(\vec r) \cdot \hat n(\vec r)$ with the surface normal vector $\hat n$ \cite{2008_Hubert_Book}.
This expression for the stray field energy can be applied to two parallel rectangular sheets of constant charge $m_a$ and $m_b$ at coordinates $(x \in [0,a], y\in[0,L], z=0)$ and $(x \in [x_0, x_0+b], y\in[0,L], z=-z_0)$. For a constant surface charge, the integration amounts to integrating $1/r$ twice with respect to each $x^\prime$ and $y^\prime$. Using appropriate choices for the integration constants this leads to
\begin{eqnarray}
 F_{220}(x,y,z)
 &=& \frac{1}{2} \left[ x(y^2-z^2) \, \text{atanh}\left(\frac{x}{r}\right) + y(x^2-z^2) \, \text{atanh}\left(\frac{y}{r}\right) \right]
 \nonumber \\
 &\hspace{1 cm}  -& xyz \,\text{arctan} \left(\frac{xy}{zr}\right)  +  \frac{1}{6} r(3z^2-r^2)
\end{eqnarray}
with $r =\sqrt{x^2+y^2+z^2}$ \cite{2008_Hubert_Book}.
Insertion of the appropriate boundaries gives the interaction energy of two parallel sheets as
\begin{eqnarray}
E_d^\parallel(m_a, m_b, a, b, x_0, z_0, L) %\nonumber\\
 &=& \frac{\mu_0}{2\pi} m_a m_b \left[ F_{220}(a-x_0-b, 0, z_0) - F_{220}(-x_0-b, 0, z_0) \right. \nonumber \\
 &\hspace{1 cm}  -& \left.  F_{220}(a-x_0, 0, z_0) + F_{220}(-x_0, 0, z_0) \right. \nonumber \\
 &\hspace{1 cm}  -& \left.  F_{220}(a-x_0 -b, L, z_0) +  F_{220}(-x_0 -b, L, z_0) \right. \nonumber \\
 &\hspace{1 cm}  +& \left. F_{220}(a-x_0, L, z_0) - F_{220}(-x_0, L, z_0)  \right] \nonumber \\
 &=& m_a m_b  \tilde E_d^{\parallel}(a, b, x_0, z_0, L).
\end{eqnarray}
The total stray-field energy can be decomposed in sums of pairwise interactions between such rectangular sheets of constant charge for the domain configuration in Extended Data Fig.~\ref{fig:EDI6}\,\textbf{a}, since the surface charge is piecewise constant on rectangular patches on the sample surface. The domain walls are not charged because the magnetization along the $y$ direction is small and neglected here. The energy contribution of sheets perpendicular to each other cancel due to symmetry, such that no mixed terms of the form $m^x m^z$ arise.

In addition to the stray-field energy, the domain-wall energy must be taken into account. The total energy cost for domain walls in the sample is assumed to be
\begin{equation}
E_{\rm dw} = \sigma_{\rm dw} A_{\rm dw} N_{\rm dw} \frac{|\vec m_1-\vec m_2|^2}{f^2}
\end{equation}
where $A_{\rm dw}=L^2$ is the area of a domain wall and $N_{\rm dw}=2N-1$ the number of domain walls. \cite{2009_Biltmo_EPL} The denominator $f=g\mu_B M/V$ compensates the units of the magnetization density, where $M$ is the number of lattice sites, $V= V_{\rm uc} M/4$ the sample volume and $V_{\rm uc}=2.5\cdot10^{-28}$\,m$^3$ the unit cell volume, which contains $4$ Ho$^{3+}$ ions \cite{1996_Bitko_PRL}. The energy density of domain walls in LiHoF$_4$ was estimated from a Monte Carlo study\cite{2009_Biltmo_EPL} as $\sigma_{\rm dw} =5.9 \cdot 10^{-7}$\,J/m$^2$.

The domain energy, i.e. the sum of $E_d$ and $E_{\rm dw}$, is bilinear in the magnetization components. It contains pieces $m^x_1 m^x_2$ and $m^z_1 m^z_2$ which can be interpreted as interaction between the domain magnetizations which is antiferromagnetic in character, as neighboring domains favor antiparallel alignment to minimize the stray-field energy.

%\bigskip

%%%%%%%%%%%%%%%%%%%%%%%%%%%%%%%%%%
\section{Combined Hamiltonian and mean-field approximation}
\label{sec:combined}

We now proceed to combine the microscopic interactions and the domain energy into a joint mean-field description. To this end, we express the domain energy in terms of microscopic moments. Since the nuclear $g$ factor is tiny compared to the electronic $g$ factor, neglecting the contribution of nuclear spins to the magnetization within each domain is a reasonable approximation. Therefore we relate the magnetization density to the electronic moments as
\begin{equation}
\vec m = g \mu_B \vec J M/V\,.
\end{equation}
With our assumption of a homogeneous magnetization within each domain, the total domain energy can be re-written in terms of the expectation values of electronic moments $\vec{\bar J}_{1,2}$ in the domains $1$ and $2$:
\begin{equation}
\label{eq:edom}
E_{\rm dom} = M \sum_{\alpha} \left( c^\alpha_1 \bar J^\alpha_1 \bar J^\alpha_1 + c^\alpha_2 \bar J^\alpha_2 \bar J^\alpha_2 + c^\alpha_{12} \bar J^\alpha_1 \bar J^\alpha_2 \right) ,
\end{equation}
with the parameters $c^\alpha$ containing the potential-energy contributions of the surface charges caused by the magnetization component $\alpha={x,y,z}$ as well as the domain-wall energies. The symmetry of the domain configuration, Extended Data Fig.~\ref{fig:EDI6}\,\textbf{a}, dictates $c^x_a=c^z_a$ for $a \in \{1,2,12\}$. Moreover, the parameters $c^y$ are not needed since $m^y$ is small and can be neglected.
Inserting the expressions derived above, we find the effective couplings as
\begin{widetext}
\begin{align}
c^\alpha_n &=  \sigma_{\rm dw} \frac{V_{uc}}{4 k_B D} + \frac{4 g^2 \mu_B^2}{L^3 V_{uc} k_B} \sum_{i=0}^{N-1} \sum_{j=0}^{N-1}  \left[ \tilde E_d^\parallel(D_n,D_n,(i-j)D,0,L) - \tilde E_d^\parallel(D_n,D_n,(i-j)D,L,L)  \right] \quad (n=1,2),
  \nonumber \\
c^\alpha_{12} &= - \sigma_{\rm dw} \frac{V_{uc}}{2 k_B D} + \frac{4 g^2 \mu_B^2}{L^3 V_{uc} k_B} \sum_{i=0}^{N-1} \sum_{j=0}^{N-1} \left[  \tilde E_d^\parallel(D_1,D_2,D_1+(i-j)D,0,L)- \tilde E_d^\parallel(D_1,D_2,D_1+(i-j)D,L,L)
  \right.  \nonumber \\ & \hspace{1cm} \left.
 + \tilde E_d^\parallel(D_1,D_2,-D_1+(i-j)D,0,L)- \tilde E_d^\parallel(D_1,D_2,-D_1+(i-j)D,L,L)\right]
 \label{eq:effcouplings}
\end{align}
\end{widetext}
for $\alpha=x,z$. For equal-sized domains, $D_1=D_2$ or equivalently $v=1/2$, we have $c^\alpha_1=c^\alpha_2$. Moreover, $c^\alpha_{12}$ is symmetric under the exchange of $D_1$ with $D_2$.
The sum of the coefficients $c_s = c_{\alpha 1} + c_{\alpha 2} + c_{\alpha 12}$ gives the domain energy of the single-domain state, $E_{\text{dom}}=M c_s \vec{\bar J}^2$. It is therefore independent of the domain ratio $v$.

Numerical evaluation for $v=1/2$, $N=200$ (and all other parameters as given above) yields $c^\alpha_1 = c^\alpha_2 \approx 6.5$\,mK, $c^\alpha_{12} \approx 13$\,mK, resulting in $c_s \approx 26$\,mK. The coefficients vary smoothly and monotonically as a function of $v$. For instance, at $v=1/4$ we have $c^\alpha_1 \approx 14.5$\,mK, $c^\alpha_2 \approx 1.5$\,mK and $c^\alpha_{12} \approx 10$\,mK. In practice, we fit the dependence on $v$ by a fifth-order polynomial in $v$ to save computation time when optimizing $v$.
The fact that $c^\alpha_{12}>0$ reflects that the effective interaction between domains due to the minimization of the stray-field energy is antiferromagnetic. Similarly, $c^\alpha_{1,2}>0$ within each domain, implying that the stray field competes with the ferromagnetic interaction between the individual moments.
Further discussion on the choice of $N$ and its influence on the resulting energetics can be found in Section~\ref{sec:energy}; numerical results are shown for $N=200$ unless noted otherwise.

We are now in the position to combine the microscopic Hamiltonian Eq.\,\eqref{eq:hmic1} with the domain energy Eq.\,\eqref{eq:edom}. Introducing domain-dependent mean fields, $\vec{\bar J}_{1,2}$, and neglecting the weak microscopic interaction across domain walls, we obtain from $H_{\rm mic}$ two independent mean-field problems of the form \eqref{eq:hmic2}, one for each domain. These two problems get modified and coupled by the domain energy. The condition of minimizing the total free energy can then be cast into two coupled mean-field Hamiltonians
\begin{eqnarray}
\label{eq:hmf1}
H^\text{MF}_1
&=& \left(-\frac{n}{2}K + \frac{c^x_1}{1-v}\right) \left( 2 \bar J^x_1 J^x_1 -  ({\bar J}_1^x)^2 \right) \nonumber \\
&+& \frac{c^x_{12}}{1-v} \left(\bar J^x_2 J^x_1  - \frac{1}{2} \bar J^x_1 \bar J^x_2 \right) \nonumber \\
&+& \left(-\frac{n}{2}K + \frac{c^z_1}{1-v}\right)\left( 2 \bar J^z_1 J^z_1 - (\bar J^z_1)^2 \right) \nonumber \\
&+& \frac{c^z_{12}}{1-v} \left( \bar J^z_2  J^z_1 - \frac{1}{2} \bar J^z_1 \bar J^z_2 \right) \nonumber \\
&-& \frac{n}{2}K \left( 2 \bar J^y_1  J^y_1 - (\bar J^y_1)^2 \right) +  H_{\text{ion}}(\hat {\vec {J}}_1),
\end{eqnarray}
and
\begin{eqnarray}
\label{eq:hmf2}
H^\text{MF}_2
&=& \left(-\frac{n}{2}K + \frac{c^x_2}{v}\right) \left( 2 \bar J^x_2  J^x_2 - (\bar J^x_2)^2 \right) \nonumber \\
&+& \frac{c^x_3}{v} \left( \bar J^x_1 J^x_2 - \frac{1}{2} \bar J^x_1\bar J^x_2 \right) \nonumber \\
&+& \left(-\frac{n}{2}K + \frac{c^z_2}{v}\right)\left( 2 \bar J^z_2 J^z_2 - (\bar J^z_2)^2 \right) \nonumber \\
&+& \frac{c^z_{12}}{v} \left( \bar J^z_1 J^z_2 - \frac{1}{2} \bar J^z_1 \bar J^z_2 \right) \nonumber \\
&-&  \frac{n}{2}K \left( 2 \bar J^y_2 J^y_2 - (\bar J^y_2)^2 \right) + H_{\text{ion}}(\hat {\vec {J}}_2),
\end{eqnarray}
with the self-consistency conditions
\begin{equation}
\label{eq:hmf3}
\vec{\bar J}_n = \langle \vec J_n \rangle
\end{equation}
where the expectation value is taken with respect to $H^\text{MF}_n$. This is apparently equivalent to promoting $E_{\rm dom}$ Eq.\,\eqref{eq:edom} to a bilinear Hamiltonian and decoupling its interaction terms in a  mean-field fashion, resulting in the terms $\bar J^\alpha_1 J^\alpha_1$ etc. The total energy then reads
\begin{equation}
E^{\rm MF}_{\rm tot} = M (1-v) \langle H^\text{MF}_1 \rangle  + M v \langle H^\text{MF}_2 \rangle\,.
\end{equation}
The self-consistency equations \eqref{eq:hmf3} are solved iteratively. In each step, the Hamiltonians $H^\text{MF}_n$ are solved exactly via direct diagonalization in the $17\times 8$-dimensional local Hilbert space. Depending on the initial conditions, in particular the relative sign of $\bar J^z_1$ and $\bar J^z_2$, one either obtains a single-domain or a multi-domain solution. Comparison of the total free energy $F$ yields the stable state.

For each set of parameters $(B, T, \phi)$ the optimum domain ratio $v$ is obtained by minimizing the free energy $F$ with respect to $v$. Since $F(v)$ is parabolic and smooth, Extended Data Fig.~\ref{fig:EDI6}\,\textbf{b}, the optimum value $v^*$ can be accurately found by means of a parabolic fit through a relatively small number of test points.

It is important to emphasize that Eqs.~(\ref{eq:hmf1}-\ref{eq:hmf3}) constitute a combined and consistent description of the microscopic interactions, the mesoscopic domain energy terms, and the backaction of the stray-field physics on the microscopic expectation values.

%%%%%%%%%%%%%%%%%%%%%%%%%%%%%%%%%%

\section{Importance of the domain shape and size}
\label{sec:energy}

For a numerical assessment of our model we calculated at first the dependence of the domain energy $E_{\rm dom}$ \eqref{eq:edom} on the number of domains. On the one hand this provided information on the relative importance of the stray fields as compared to the contributions by domain walls. On the other hand this served to optimize the computational effort and allowed to choose a suitable domain number and thus domain size for meaningful results.

Shown in Extended Data Fig.\,\ref{fig:EDI9} is the domain energy per site, $E_{\rm dom}/M$, as a function of the number of domains, $N$, for $N$ between 1 and $10^3$. The calculation assumes a sample size of $5\,{\rm mm}$ such that $N$ translates into domain sizes between mm and several ${\rm \mu m}$. At zero magnetic field, the latter correspond to typical domain sizes observed experimentally in {\lhf} \cite{1975_Battison_JPC, 1988_Pommier_JdP, 1989_Meyer_EOMOMatAppl, MAJorba, 2014_Karci_RSI} as well as related compounds such as LiTbF$_4$ \cite{1990_Koetzler_PRL}.

As shown in Extended Data Fig.\,\ref{fig:EDI9}\,\textbf{a} at zero magnetic field the domain-wall energy is tiny in comparison to the stray-field energy and proportional to the number of domains $N$, since the number of domain walls is proportional to $N$. In comparison, the stray-field energy varies as $1/N$. This can be understood as follows: The main contribution of the stray fields is from within each domain, since any contribution beyond the size of the domain is reduced by the staggered alignment of first-neighbor and second-neighbor domains. This contribution to the stray-field energy scales as $q^2$, where $q \propto 1/N$ represents the magnetic surface charge of each domain. Summing over all domains one finds that the energy of the stray fields scales as $N \times (1/N)^2 = 1/N$.

The competition between stray-field and domain-wall energies determines the optimal domain size, i.e. the optimal value of $N$, as hinted in Extended Data Fig.\,\ref{fig:EDI9}\,\textbf{b}. The computational effort to determine the stray field energy of each domain scales as $N^2$, such that the minimal domain energy, located at $N>1000$, is difficult to access.

In the presence of an applied magnetic field parallel to the $x$-axis a substantial homogeneous magnetization is added. Because its additional contribution to the stray-field energy is independent of the staggered arrangement of the domains, it is independent of $N$, in agreement with our results obtained numerically, see Extended Data Fig.\,\ref{fig:EDI9}\,\textbf{a} and \ref{fig:EDI9}\,\textbf{b}. At the same the energy of the domain walls reduces very weakly. Taken together, the results for the domain energy become less sensitive to the details of the domains with increasing field.

Our simulations were carried out for sheet-like domains (cf. Extended Data Fig.\,\ref{fig:EDI6}). It is instructive to hypothetically assume that the domain-wall energy were larger by a factor of $100$ than they are really. Then the optimal $N$ were around $200$, see Extended Data Fig.\,\ref{fig:EDI9}\,\textbf{c}. It is clear that minimum gets shallower as the field increases, so especially near the critical field the results are not sensitive to the precise value of $N$. We have therefore performed all simulations for $N = 200$.

%%%%%%%%%%%%%%%%%%%%%%%%%%%%%%%%%%
\section{Numerical results and discussion}
\label{sec:numerical}

%%%%%%%%%%%%%%%%%%%%%%%%%%%%%%%%%%
\subsection{Phases and phase diagram}
\label{sec:num-pd}

In zero field, we find the standard mean-field behavior. At temperatures $T>T_c$ the mean fields vanish, corresponding to a paramagnet at finite temperature. For $T<T_c$ solutions with finite mean fields are found, with a multi-domain state having lower free energy than a single-domain state -- this simply reflects domain formation in the ferromagnet.
In contrast, with a large magnetic field applied, a field-polarized single-domain state is most favorable. This implies that there is a sharp field-driven multi-domain to single-domain transition for any temperature $T<T_c$ and any field direction. For $\phi=0$ this domain transition at $B_c$ coincides with the microscopic transition involving broken Ising symmetry. The multi-domain state for $B<B_c$ has $\bar J^z_{1,2}\neq 0$, while $\bar J^z_{1,2}=0$ for $B>B_c$.
The resulting phase diagram is shown in Fig.~\ref{fig:3}\,\textbf{h}. The phase boundary corresponding to the domain transition follows $B_c(\phi\!=\!0)-B_c \propto \phi^2$ at small $\phi$ in agreement with experiment.

Our calculation, which correctly accounts for the stray fields, automatically incorporates demagnetization effects. For instance, the uniform susceptibility $\chi^{zz}$, which one would associate with the order-parameter susceptibility of the ferromagnet, does \emph{not} diverge at the transverse-field transition. Moreover, the variability of the domain ratio $v$ implies that $\chi^{zz}$ is large (and essentially constant) throughout the entire multi-domain phase, Fig.~\ref{fig:3}\,\textbf{h}, because the system responds to a change in $B^z$ by a change in $v$, i.e., by shifting domain walls\cite{1975_Cooke_JPC, 1978_Beauvillain_PRB}.

Numerical results for the domain ratio are shown in Fig.~\ref{fig:4}\,\textbf{b} and Extended Data Fig.~\ref{fig:EDI6}\,\textbf{c}. For transverse field, $\phi=0$, the domain ratio is $v^*=1/2$ by symmetry, while for tilted fields $v^*$ is determined by a competition between the Zeeman and stray-field energies. When approaching the transition field from below for $\phi\neq 0$, the minority domains, whose $z$ component of the magnetization is antiparallel to the $z$ component of the field, are squeezed out, i.e., $v^*$ approaches zero continuously. The properties of the domain transition evolve continuously from small $\phi$ to large $\phi$. For $\phi=90^\circ$ $B_c$ can be identified with the coercive field.

The suppression of the minority domains when crossing $B_c$ at finite $\phi$ implies that the magnetization along the field displays a sharp kink at $B_c$, Fig.~\ref{fig:4}\,\textbf{c}, and that $\chi^{zz}$ drops discontinuously, Fig.~\ref{fig:4}\,\textbf{d}, because the high-field state is devoid of moveable domain walls and displays the microscopic response. The field and angle dependence of $\chi^{zz}$ in Fig.~\ref{fig:4}\,\textbf{d} is in excellent agreement with the experimental data for $\chi'$ in Fig.~\ref{fig:2}, with the only difference that the observed domain-wall freezing (which may be subject to defect pinning) is absent in the theoretical calculation. Note that the calculation determines the static susceptibility, whereas the experiment probes the ac susceptibility at 511\,Hz, but the frequency dependence of $\chi'$ is vanishingly small for $B>B_c$ and $T>T_c$, i.e., far from the regime dominated by magnetic domains, as shown in Extended Data Fig.\,\ref{fig:EDI3}.
Parenthetically, we mention that magnetic saturation may only be reached for fields much larger than the CEF energies, i.e., for $B \gtrsim 250$\,T.

It is interesting to note, that the microscopic properties of the model taking into account the effect of stray fields, exhibits a crossover akin that observed in the purely microscopic model at $B^*$ (cf. Fig.\,\ref{fig:3}\,\textbf{d}). However, the crossover may only be seen in the regime of the phase diagram not dominated by magnetic domains, i.e., for $B>B_c$ and $T>Tc$. For the sake of clarity we did not mark the location of this cross-over in the phase diagrams shown in Figs.\,\ref{fig:3}\,\textbf{h} and \ref{fig:4}\,\textbf{e}.

The $B$-$T$ phase diagram calculated theoretically for different angles $\phi$ is shown in Fig.~\ref{fig:4}\,\textbf{e} (recall also Fig.\,\ref{fig:3}\,\textbf{h} for the zero temperature limit). The quantitative agreement between theory and experiment is remarkably good, in particular given the simplicity of our approximations.
We recall that we have ignored the dipolar character of the interaction and that we have made fairly crude assumptions concerning the domain shape and homogeneity. Concerning the latter we believe that the spherical form of the sample measured in this study is advantageous, because it leads to uniform internal fields. We note that the experimental transition at finite $\phi$ appears to be slightly broadened as shown Fig.\,\ref{fig:2}; this is likely due to tiny residual inhomogeneities in the magnetization and domain distribution across the sample.

%%%%%%%%%%%%%%%%%%%%%%%%%%%%%%%%%%
\subsection{Interplay of non-Kramers moments and hyperfine coupling}
\label{sec:num-kram}

We finally discuss the surprising sensitivity of the transition field $B_c$ to small tilt angles and the disappearance of the inflection point in the phase boundary as a function of temperature for $\phi\gtrsim 5^\circ$ (cf. Extended Data Fig.~\ref{fig:EDI5}). Both are related to a combination of non-Kramers physics and strong hyperfine coupling.
First, a key property of the CEF levels of the Ho ions concerns that a finite expectation value $\langle J^x \rangle$ cannot be realized by a superposition of the non-Kramers doublet states, but only via interaction with higher CEF states. As a consequence, the magnitude of the magnetic moment $|J| = \sqrt{\langle \vec{J}^2 \rangle}$ is not constant as a function of $B$, but has a pronounced minimum near $B_c$ because the field-induced $x$ component grows quadratically only. In tilted fields this minimum of $\sqrt{\langle \vec{J}^2 \rangle}$ is shallower because $\langle J^z \rangle$ decreases more slowly and remains finite due to the longitudinal field component, Extended Data Fig.~\ref{fig:EDI5}\,\textbf{b}.
Second, the hyperfine coupling $A$ energetically prefers large electronic moments $|J|$, i.e., more hyperfine energy can be gained by anti-alignment of nuclear spins if $|J|$ is larger. Hence, the hyperfine coupling stabilizes the ferromagnetic with respect to the paramagnetic phase for temperatures $T\lesssim 0.5$\,K. This leads to the low-temperature hump in the phase boundary \cite{1996_Bitko_PRL}, see Fig.~\ref{fig:2}\,\textbf{e} and Fig.~\ref{fig:4}\,\textbf{e} and Extended Data Fig.\,\ref{fig:EDI7}\,\textbf{a}.
Third, with increasing tilt angle $\phi$, the variation of $|J|$ gets less pronounced, such that the relative energy gain of the ordered phase decreases and the critical field is only mildly enhanced by hyperfine effects. This explains both the absence of the hump of the phase-boundary for $\phi\gtrsim 5^\circ$ and the strong variation of $B_c$ for $\phi<5^\circ$ as shown in Figs.~\ref{fig:2}\,\textbf{h}, \ref{fig:2}\,\textbf{i}, \ref{fig:4}\,\textbf{e}, and Extended Data Fig.\,\ref{fig:EDI7}\,\textbf{a}. In short, tilted fields rapidly eliminate the non-Kramers variation of $|J|$ and hence the hyperfine-induced enhancement of $B_c$.

For comparison, we repeated the same calculation without the CEF terms, where the ordered moments are standard $J=8$ spins, implementing the Ising anisotropy via a single-ion term of the form $-D {J^z}^2$ with $D=0.215$\,K. While all qualitative properties of the transitions remain unchanged,  the critical field is now essentially independent of the hyperfine coupling, Extended Data Fig.~\ref{fig:EDI5}\,\textbf{c}, because the magnetic moment $|J|$ is nearly constant across the phase transition, with only tiny variations due to entanglement with the nuclear spins, see Fig.~\ref{fig:EDI5}\,\textbf{d}.
We conclude that the non-Kramers physics magnifies the effects of hyperfine coupling, which proves to be very helpful for disentangling the microscopic from the mesoscale effects.

%%%%%%%%%%%%%%%%%%%%%%%%%%%%%%%%%%
%%%%%%%%%%%%%%%%%%%%%%%%%%%%%%%%%%
\newpage
\section*{Supplementary References}
%\bibliography{LiHoF4-v2}

%%%%%%%%%%%%%%%%%%%%%%%%%%%%%%%%%%

%%%%%%%%%%%%%%%%%%%%%%%%%%%%%%%%%%
%%%%%%%%%%%%%%%%%%%%%%%%%%%%%%%%%%

\end{document}